\documentclass[fleqn,usenatbib]{mnras}

\usepackage{newtxtext,newtxmath}

\usepackage[T1]{fontenc}

\DeclareRobustCommand{\VAN}[3]{#2}
\let\VANthebibliography\thebibliography
\def\thebibliography{\DeclareRobustCommand{\VAN}[3]{##3}\VANthebibliography}

\usepackage{graphicx}	
\usepackage{amsmath}	
\usepackage{fontspec}
\usepackage{bm}

\title[\texttt{Aletheia}: Emulating $P(k)$ with evolution mapping]{\texttt{Aletheia}: Emulating the non-linear matter power spectrum in the context of evolution mapping}

\author[A. G. S\'anchez et al.]{
Ariel G. S\'anchez,$^{1,2}$\thanks{E-mail: arielsan@mpe.mpg.de}
Andr\'es N. Ruiz,$^{3,4}$
Facundo Rodriguez,$^{3,4}$
Carlos Correa,$^{1}$
Andrea Fiorilli,$^{1}$\newauthor
Matteo Esposito,$^{1}$
Jenny Gonzalez-Jara,$^{5}$
Nelson D. Padilla,$^{3,4}$
Alejandro P\'erez-Fern\'andez,$^{1}$\newauthor
and Sofia Contarini$^{1}$
\\
$^{1}$Max-Planck-Institut f\"ur extraterrestrische Physik, Postfach 1312, Giessenbachstr., 85748 Garching, Germany\\
$^{2}$Universit\"as-Sternwarte M\"uchen,  Fakult\"at f\"ur Physik, Ludwig- Maximilians-Universit\"at M\"unchen, Scheinerstrasse 1, 81679 M\"uchen, Germany\\
$^{3}$Instituto de Astronomía Teórica y Experimental, CONICET-UNC, Laprida 854, X5000BGR, Córdoba, Argentina\\
$^{4}$Observatorio Astronómico, Universidad Nacional de Córdoba, Laprida 854, X5000BGR, Córdoba, Argentina\\
$^{5}$Instituto de Astrofísica, Pontificia Universidad Católica de Chile, Av. Vicu\~na Mackenna 4860, Santiago, Chile 
}

\date{Accepted XXX. Received YYY; in original form ZZZ}

\pubyear{\the\year{}}

\begin{document}
\label{firstpage}
\pagerange{\pageref{firstpage}--\pageref{lastpage}}
\maketitle

\begin{abstract}
We present {\tt Aletheia}, a new emulator of the non-linear matter power spectrum, $P(k)$, built upon the evolution mapping framework. This framework addresses the limitations of traditional emulation by focusing on $h$-independent cosmological parameters, which can be separated into those defining the linear power spectrum shape ($\bm{\Theta}_{\mathrm{s}}$) and those affecting only its amplitude evolution ($\bm{\Theta}_{\mathrm{e}}$). The combined impact of evolution parameters and redshift is compressed into a single amplitude parameter, $\sigma_{12}$. {\tt Aletheia} uses a two-stage Gaussian Process emulation: a primary emulator predicts the non-linear boost factor as a function of ($\bm{\Theta}_{\mathrm{s}}$) and $\sigma_{12}$ for fixed evolution parameters, while a second one applies a small linear correction based on the integrated growth history. The emulator is trained on shape parameters spanning $\pm$5$\sigma$ of Planck constraints and a wide clustering range $0.2 < \sigma_{12} < 1.0$, providing predictions for $0.006\,{\rm Mpc}^{-1} < k < 2\,{\rm Mpc}^{-1}$. We validate {\tt Aletheia} against N-body simulations, demonstrating sub-percent accuracy. When tested on a suite of dynamic dark energy models, the full emulator's predictions show a variance of approximately 0.2\%, a factor of five smaller than that of the state-of-the-art {\tt EuclidEmulator2} (around 1\% variance). Furthermore, {\tt Aletheia} maintains sub-percent accuracy for the best-fit dynamic dark energy cosmology from recent DESI data, a model whose parameters lie outside the training ranges of most conventional emulators. This demonstrates the power of the evolution mapping approach, providing a robust and extensible tool for precision cosmology.
\end{abstract}




\section{Introduction}
\label{sec:intro}

The large-scale structure (LSS) of the Universe, as traced by the spatial
distribution of galaxies, provides one of the most powerful avenues for
constraining cosmological models and understanding the fundamental nature of
gravity, dark matter, and dark energy. Analyses of LSS, particularly through
measurements of baryon acoustic oscillations (BAO) and redshift-space
distortions (RSD), have become cornerstone techniques in modern cosmology
\citep[e.g.,][]{Eisenstein2005, Cole2005, Sanchez2006, Anderson2014, Alam2017,
Alam2021}. The recent results from the Dark Energy Spectroscopic Instrument
\citep[DESI,][]{DESI2024_FS1, DESI2025_BAO2, Adame2025_DESIBAO1}, have further
underscored the constraining power of these methods. Together with the
\textit{Euclid} mission \citep{Euclid_overview}, these datasets promise to usher
in an era of unprecedented precision. These advancements in galaxy surveys are
complemented by a wealth of new data from several cosmological probes, including
cosmic microwave background (CMB) measurements from the \textit{Planck}
satellite \citep{Planck2018, Planck2024} and ground-based experiments like the
Atacama Cosmology Telescope \citep[ACT,][]{Thibaut2025_ACT}, type Ia Supernovae
(SN) compilations \citep[e.g.,][]{Scolnic2018_Pantheon,
Scolnic2022_PantheonPlus, Brout2022_PantheonPlus}, and weak lensing shear data
from surveys such as the Dark Energy Survey \citep[DES,][]{Abbott2022}, the
Kilo-Degree Survey \citep[KiDS,][]{Wright2025}, and the Hyper Suprime-Cam Subaru
Strategic Program \citep[HSC,][]{Dalal2023, Li2023}, marking a truly data-rich
era in cosmology.

This influx of high-quality data is beginning to reveal intriguing, albeit
tentative, hints of physics beyond the standard Lambda cold dark matter
($\Lambda$CDM) model. For instance, the combination of DESI data with Planck and
SN observations shows a preference for dynamic dark energy models over a
cosmological constant \citep{Adame2025_DESIBAO1, DESI2025_BAO2}. To rigorously
test such claims and fully exploit the potential of forthcoming datasets, we
require analysis tools that are not only highly accurate but also robust across
a wide range of cosmological scenarios.

At their core, the current state-of-the-art LSS analyses, including those from
DESI, build upon methodologies developed for previous surveys \citep[e.g.,][]{Sanchez2017b, Grieb2017, Troster2020, Damico2020, Ivanov2020}. These
rely on perturbation theory to model the clustering of galaxies in the mildly
non-linear regime. However, these recipes inherently struggle as the analyses
are pushed to smaller scales, where non-linear evolution becomes dominant and
perturbative approaches become invalid. 

An alternative is to use N-body simulations directly as theoretical predictions.
While ideal for capturing non-linear evolution, the computational expense of
running such simulations for every point in a vast cosmological parameter space
is prohibitive for most analyses. Fortunately, non-linear clustering statistics,
like the matter power spectrum, $P(k)$, typically exhibit a smooth dependence on
cosmological parameters. This characteristic enables the development of
emulators: sophisticated interpolation schemes trained on a set of N-body
simulations. Once calibrated, emulators can deliver rapid and accurate
predictions for non-linear quantities \citep[e.g.,][]{coyote2010, Heitman2016,
Garrison2018, AemulusI, Bocquet2020, Euclidemulator, EuclidEmulator2, Angulo2021_BACCO,
Yang2025}. 

Despite their utility, emulators are inherently limited by the parameter space
and redshifts sampled during their calibration. The high computational cost of
the underlying simulations means that the number of training nodes must be kept
manageable. The vast parameter space of possible cosmological models means that
current emulators are often focused on narrow regions around a fiducial
$\Lambda$CDM cosmology to maintain accuracy, limiting their applicability to
explorations of more exotic models or broader parameter variations.

These challenges are exacerbated by the very language traditionally used to
describe cosmic structure, which is commonly based on quantities that depend
explicitly on the dimensionless Hubble parameter $h$.
\citet{Sanchez2020} highlighted the complications associated with this common
practice, which obscures the information content of different cosmological observables.

Building directly upon these insights, the evolution mapping framework
introduced in \citet{Sanchez2022} offers a new framework that reframes how we
model the non-linear Universe. Its core principle is to adopt a parametrisation
free from any explicit dependence on the dimensionless Hubble
parameter. This enables a separation of cosmological parameters into two
distinct classes: shape parameters, which determine the shape of the linear
matter power spectrum, and evolution parameters, which govern the subsequent
growth of structure. 
At the linear level, the impact of different evolution
parameters can be compressed into a single variable characterising the global
amplitude of density fluctuations. 

\citet{Sanchez2022} also outlined the practical application of evolution mapping
for designing a new class of emulators of the matter power spectrum or other
statistics. In this scheme, the emulator does not directly sample the full space
of evolution parameters, nor does it need to explicitly model the redshift
dependence of $P(k)$. Instead, by focusing on how the non-linear $P(k)$ responds
to changes in the clustering amplitude, this approach can lead to more general
predictions valid over a wider range of cosmologies and redshifts. The reduced
dimensionality of the problem leads to more accurate predictions from a given
number of training simulations.

In this paper, we present {\tt Aletheia}\footnote{From the ancient Greek
\textit{alētheia} (pronounced /aˈleː.tʰeː.a/), meaning ‘truth’ or ‘disclosure’.
In Greek mythology, Aletheia was the personification of truth.}, a new emulator
for the non-linear matter power spectrum built upon the evolution mapping
framework. {\tt Aletheia} is designed as a proof-of-concept demonstrating the
power and flexibility of this emulation strategy. Specifically, we emulate the
ratio of the non-linear $P(k)$ to its de-wiggled linear counterpart (preserving
the broadband shape but damping the BAO signal) as a function of shape parameters
and clustering amplitude for a fixed set of reference evolution parameters.
The effect of varying evolution parameters is captured by a second emulator for
the dependence of the power spectrum on the growth of the structure history. This
allows {\tt Aletheia} to efficiently make predictions for general cosmologies
while relying on simulations that provide only a single snapshot for each
sampled cosmology. We construct {\tt Aletheia} using Gaussian Process regression
trained on a suite of N-body simulations specifically designed for this
emulator. We validate its performance against independent test simulations,
including a dedicated set of dynamic dark energy models, demonstrating excellent
accuracy. {\tt Aletheia} shows good agreement with other established emulators
such as \texttt{EuclidEmulator2} \citep{EuclidEmulator2}, while offering
predictions over a wider range of cosmological models and redshifts. 

The structure of this paper is as follows. In Section~\ref{sec:evo_mapping}, we
briefly review the general idea of evolution mapping and introduce the parameter
$\tilde{x}$ used to describe deviations from the exact evolution mapping
relation. Section~\ref{sec:emulating_pk} details the design of the {\tt
Aletheia} emulator, including the parameter space sampled
(Section~\ref{ssec:emulator_design}), the cosmological simulations used for
training (Section~\ref{ssec:simulations}), the Gaussian Process emulation
methodology (Section~\ref{ssec:gp_emulation}),  and the implementation of a
high-$k$ resolution correction necessary for the final emulator product
(Section~\ref{sec:resolution}). We present a thorough validation of {\tt
Aletheia} in Section~\ref{sec:validation}, assessing its individual components
(Section~\ref{ssec:validation_components}), its performance for dynamic dark
energy cosmologies compared against other publicly available emulators
(Section~\ref{ssec:validation_full_emu}), and the impact of the high-$k$
resolution correction (Section~\ref{ssec:validation_resolution}). Finally, we
summarise our findings and discuss future prospects in
Section~\ref{sec:conclusions}.

\begin{figure*}
\includegraphics[width=0.95\textwidth]{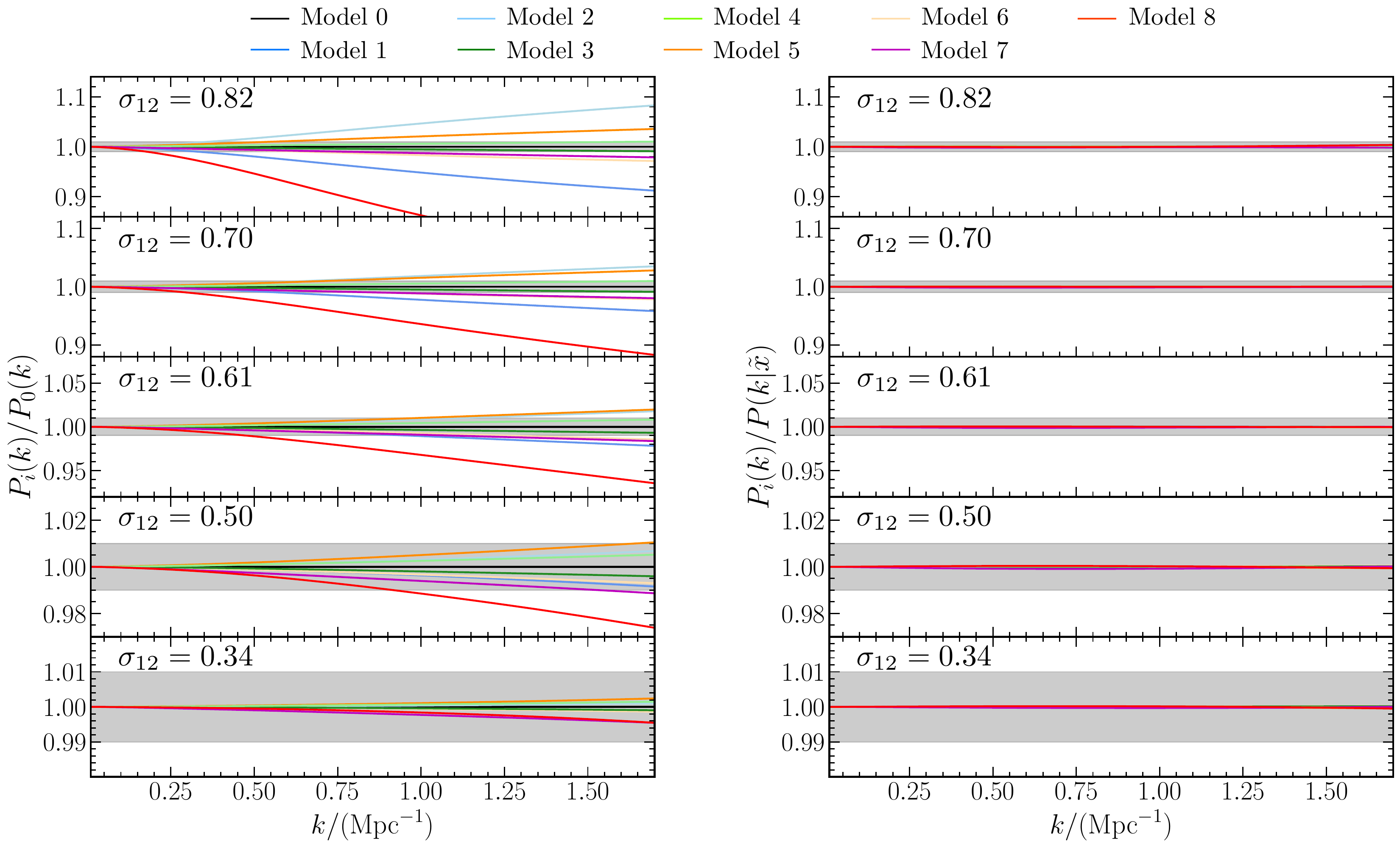}
    \caption{
    The left panel shows the ratio of non-linear power spectra for a set of simulations with identical shape parameters but widely varying evolution parameters, all evaluated at redshifts that correspond to the same values of $\sigma_{12}$. The specific cosmological parameters for each model are detailed in Table 2 of \citet{Esposito2024_VelEvoMap}. The right panel shows the ratio of the true $P(k)$ to the reference prediction (using $\tilde{x}_0$) and the resulting derivative $\partial R / \partial \tilde{x}$ measured from the simulations, illustrating the accuracy of the first-order Taylor expansion in equation~(\ref{eq:pk_taylor_xtilde}).}
    \label{fig:ratios_pk_Aletheia}        
\end{figure*}

\section{The evolution mapping framework}
\label{sec:evo_mapping}

\subsection{The evolution mapping principle in the linear regime}
\label{ssec:evo_map_linear}

The framework of evolution mapping introduced by \citet{Sanchez2022} provides a powerful way to understand and model the non-linear evolution of the density
field across a wide range of cosmological models. This approach focuses on a set of cosmological parameters free from any explicit dependency on the dimensionless Hubble parameter, $h$. It classifies them according to their impact on the linear matter power spectrum, $P_{\mathrm{L}}(k)$, into two categories: 
\begin{itemize}
    \item Shape parameters, $\Theta_{\mathrm{s}}$, which define the shape of $P_{\mathrm{L}}(k)$. 
    Examples of these parameters are the physical baryon density, $\omega_{\mathrm{b}}$, the physical 
    cold dark matter density, $\omega_{\mathrm{c}}$, and the primordial scalar spectral index, 
    $n_{\mathrm{s}}$. 
    \item Evolution parameters, $\Theta_{\mathrm{e}}$, which only affect the amplitude of 
    $P_{\mathrm{L}}(k)$ at any given redshift $z$ through their impact on the growth rate of 
    cosmic structures. This set includes the physical density parameters of dark energy, 
    $\omega_{\rm DE}$, and curvature, $\omega_{\rm K}$, the primordial scalar amplitude, $A_{\rm s}$, as well as the parameters describing 
    the equation of state of dark energy, such as  $w_0$ and  $w_a$ \citep{Chevallier2001, Linder2003}.
\end{itemize}
This classification is useful as, at the linear level, the impact of all evolution parameters and $z$ follows a perfect degeneracy and can therefore be
fully described by a single quantity characterising the amplitude of $P_{\rm L}(k)$. It is common to describe the amplitude of density fluctuations in terms of the RMS of the linearly evolved mass contrast in spheres of radius
$R=8\,h^{-1}\mathrm{Mpc}$, which we denote as $\sigma_{8/h}$ to emphasise its
explicit dependence on $h$. This dependency
makes $\sigma_{8/h}$ unsuitable for cleanly describing the degeneracy involving evolution parameters. Since $h$ is given by the sum of all physical energy contributions as
\begin{equation}
h^2 = \sum_i \omega_i,
\label{eq:hubble}
\end{equation} 
it represents a combination of both shape and evolution parameters. Therefore, its inclusion in $\sigma_{8/h}$ obscures the fundamental distinction between these parameter classes and hides the degeneracy. \citet{Sanchez2020}
showed that this issue can be avoided by using a fixed reference scale in Mpc. 
A convenient choice is $\sigma_{12}$, introduced by \cite{Sanchez2020} as the RMS of linear density fluctuations smoothed over spheres of radius $12\,{\rm Mpc}$, as it provides a value
comparable to $\sigma_{8/h}$ for typical cosmological parameters. While other
physical characterisations are possible, such as $\sigma(R)$ for any $R$ in
$\mathrm{Mpc}$ units, or the dimensionless power spectrum
$\Delta^2_{\mathrm{L}}(k_{\mathrm{p}})$ at a reference wavenumber
$k_{\mathrm{p}}$ in $\mathrm{Mpc}^{-1}$ units, we follow the approach of \citet{Sanchez2022} and use $\sigma_{12}$
for consistency.

The degeneracy between the evolution parameters $\Theta_{\mathrm{e}}$
and redshift $z$ can therefore be expressed as
\begin{equation}
P_{\mathrm{L}}(k|z,\bm{\Theta}_{\mathrm{s}},\bm{\Theta}_{\mathrm{e}}) = 
P_{\mathrm{L}}\left(k|\bm{\Theta}_{\mathrm{s}},\sigma_{12}\left(z,\bm{\Theta}_{\mathrm{s}},\bm{\Theta}_{\mathrm{e}}\right)\right).
\label{eq:pk_evmap_linear}
\end{equation}
This equation, referred to as the evolution mapping relation for the linear
power spectrum, implies that the time evolution of $P_{\mathrm{L}}(k)$ in models
with the same $\bm{\Theta}_{\mathrm{s}}$ but different
$\bm{\Theta}_{\mathrm{e}}$ can be mapped from one to another simply by
relabelling the redshifts that correspond to the same $\sigma_{12}$. 

The left panel of Fig.~2 in \citet{Sanchez2022} provides a clear illustration of
this principle, showing identical linear power spectra for nine cosmological models with identical shape parameters but vastly different
evolution parameters when evaluated at redshifts where their respective
$\sigma_{12}(z)$ values match.

This framework has been extended to describe statistics of the
velocity field \citep{Esposito2024_VelEvoMap} and to incorporate the impact of
massive neutrinos \citep{Pezzotta2025_NuEvoMap}, highlighting the robustness of the evolution mapping principle.

\begin{figure*} \includegraphics[width=0.95\textwidth]{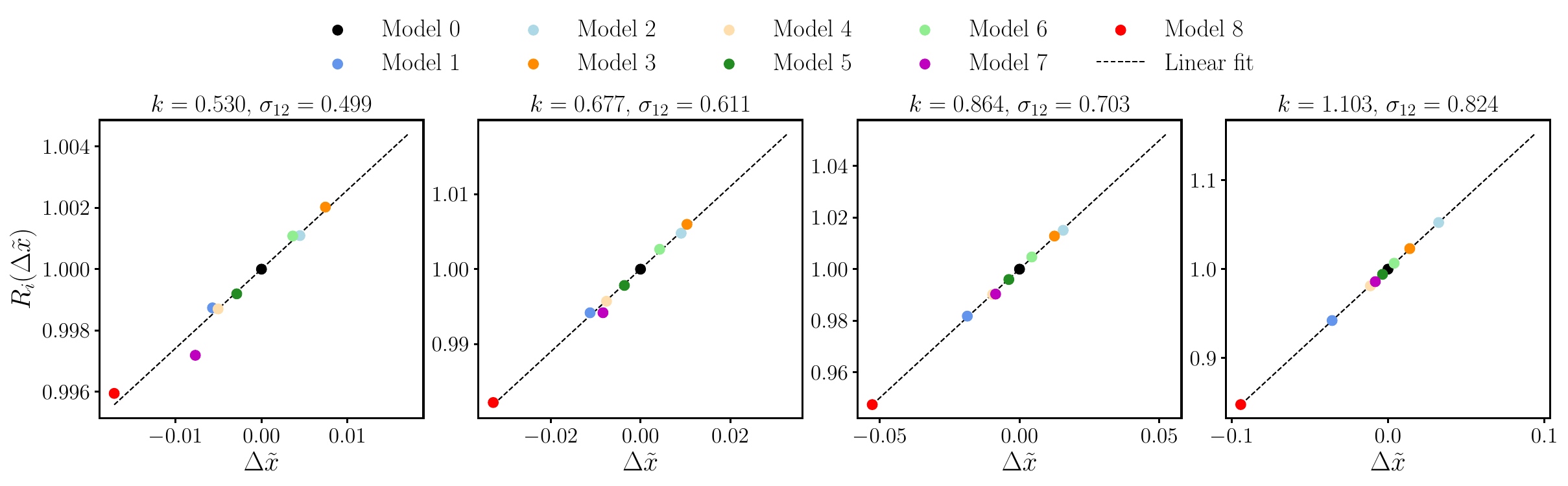}
    \caption{
    The ratio of power spectra, $R(k)$, as a function of the integrated growth history parameter $\tilde{x}$ for four choices of $k$ and $\sigma_{12}$, as indicated in the legend. The points show the measurements from the Aletheia simulations, where each colour represents a different cosmology \citep[with parameters detailed in Table 2 of ][]{Esposito2024_VelEvoMap}, all evaluated at a fixed $\sigma_{12}$. The dashed lines show the best-fitting linear relation for each case. The remarkable linearity of the response validates the first-order approximation in equation~(\ref{eq:pk_taylor_xtilde}) and allows for a direct measurement of the derivative $\partial R / \partial \tilde{x}$ from the simulations.}
\label{fig:Ri_vs_x}
\end{figure*}

\subsection{Deviations in the non-linear regime}
\label{ssec:evo_map_nonlinear}

The non-linear evolution of the matter power spectrum, $P(k)$, is predominantly determined by its linear-theory counterpart, $P_{\mathrm{L}}(k)$ \citep[e.g.,][]{Hamilton1991, Peacock1996, Smith2003}. Standard perturbation theory (SPT) and related approaches suggest that, if the perturbation theory kernels are independent of cosmology \citep[a good approximation, see e.g.,][]{Takahashi2008, Taruya2016, Garny2021}, then the non-linear $P(k)$ should be a function of $P_{\mathrm{L}}(k)$ that is independent of the specific cosmological parameters beyond those encapsulated in the power spectrum itself \citep[see, e.g,][]{Scoccimarro1998}. This is exploited by the emulator COMET, which uses evolution mapping to emulate the perturbation theory models commonly employed in present-day LSS analyses \citep{Eggemeier2023, Pezzotta2025_NuEvoMap}.
Under this assumption, the evolution mapping relation of equation~(\ref{eq:pk_evmap_linear}) would be expected to
hold even  for the fully non-linear $P(k)$.

\citet{Sanchez2022} used $P(k)$ measurements from N-body simulations to show that the evolution mapping relation provides a remarkably good description even in the non-linear regime. Indeed, cosmologies with identical shape parameters and widely different evolution parameters yield very similar non-linear structure when viewed at equal $\sigma_{12}$. However, these models exhibit small deviations in their non-linear power spectra, particularly in the deeply non-linear regime. 
This is illustrated in the left panel of Fig.~\ref{fig:ratios_pk_Aletheia} \citep[which is analogous to Fig.~3 of][]{Sanchez2022}.
These differences increase with wavenumber $k$ and with $\sigma_{12}$, although they remain at the sub-percent to few-percent level for a wide range of scales and cosmologies. 

To account for these deviations, we must identify a parameter that captures the
relevant differences in the growth history. In SPT, the primary cosmology
dependence of the solutions, beyond the linear power spectrum itself, is encoded
in the parameter combination 
\begin{equation}
x(z) = \frac{\Omega_{\mathrm{m}}(z)}{f^2(z)},
\label{eq:def_x}
\end{equation}
where $\Omega_{\mathrm{m}}(z)$ is the fractional matter density parameter and $f(z) = \mathrm{d}\ln D(z) / \mathrm{d}\ln a$ is the linear growth rate parameter \citep{Scoccimarro1998}. While the instantaneous value of $x(z)$ is important, we find that the small deviations from the evolution mapping relation are sensitive to the recent growth history.
We therefore define a new parameter, \textbf{$\tilde{x}$}, which is an
average of $x$ over the past history, using $\tau = \ln(\sigma_{12})$ as the time variable. This parameter is defined as
\begin{equation}
\tilde{x}(\tau) = \int_{-\infty}^\tau x(\tau') K(\tau-\tau'|\eta) \,d\tau',
\label{eq:def_xtilde}
\end{equation}
where $K(\tau|\eta)$ is a Gaussian kernel with a width $\eta$ that represents a characteristic memory of the non-linear evolution, indicating the $\Delta\ln(\sigma_{12})$ interval over which the past growth history significantly impacts the power spectrum at any time. We find that a fixed value of $\eta=0.12$ provides an excellent description of the deviations for all relevant cosmologies and scales considered here, so we adopt it throughout.

This new parameter, \textbf{$\tilde{x}$}, effectively describes the differences in the power spectrum due to varying structure growth histories. For cosmologies sharing the same $\bm{\Theta}_{\mathrm{s}}$ and $\sigma_{12}$, but having different values of $\tilde{x}$, we can approximate the ratio of their non-linear power spectra with a first-order Taylor expansion around a reference value $\tilde{x}_0$ as
\begin{equation}
R(k, \tilde{x}) \equiv \frac{P(k | \bm{\Theta}_{\mathrm{s}}, \sigma_{12}, \tilde{x})}{P(k | \bm{\Theta}_{\mathrm{s}}, \sigma_{12}, \tilde{x}_0)}
\approx 1 + \left. \frac{\partial R(k)}{\partial \tilde{x}} \right|_{\tilde{x}_0} (\tilde{x} - \tilde{x}_0).
\label{eq:pk_taylor_xtilde}
\end{equation}
This linear correction is sufficient to capture the bulk of the differences in $P(k)$ due to varying growth histories, forming the basis of our emulator design.

Figure~\ref{fig:Ri_vs_x} provides strong empirical support for this approach. It  shows the ratios $R(k, \tilde{x})$ measured from the Aletheia simulations of \citet{Esposito2024_VelEvoMap} for four different choices of $k$ and $\sigma_{12}$. The specific cosmological parameters for each model shown are detailed in Table 2 of \citet{Esposito2024_VelEvoMap}. The results demonstrate that the dependence of this ratio on \textbf{$\tilde{x}$} is remarkably linear. The dashed lines show the best-fitting linear relation, and their excellent agreement justifies the use of the Taylor expansion in equation~(\ref{eq:pk_taylor_xtilde}). The slope of this linear fit can be interpreted as a direct measurement of the derivative $\partial R(k) / \partial \tilde{x}$, which is the second key quantity our emulators are designed to predict. The remarkable accuracy of this first-order approximation is empirically validated in the right panel of Fig.~\ref{fig:ratios_pk_Aletheia}, where the ratio of the true $P(k)$ to the prediction using the $\tilde{x}_0$ reference model 
agrees with unity to well below the 1\% level.

\section{Emulating the matter power spectrum}
\label{sec:emulating_pk}

Building upon the evolution mapping framework described in
Section~\ref{sec:evo_mapping}, we introduce {\tt Aletheia}, a new emulator for
the non-linear matter power spectrum, $P(k)$. The design of {\tt Aletheia} aims to provide accurate and
general predictions for $P(k)$ by separating the emulation task into two
distinct components. This approach uses the fact that once the values the
shape parameters $\bm{\Theta}_{\mathrm{s}}$ and $\sigma_{12}$ are fixed, the
non-linear $P(k)$ is largely determined, with only minor corrections needed to
account for differing structure growth histories.

\subsection{Emulator design}
\label{ssec:emulator_design}

The {\tt Aletheia} emulator predicts the non-linear matter power spectrum,
$P(k|\bm{\Theta}_{\mathrm{s}}, \bm{\Theta}_{\mathrm{e}},z)$, for a target cosmology defined by its shape parameters, $\bm{\Theta}_{\mathrm{s}} = (\omega_{\mathrm{b}},
\omega_{\mathrm{c}}, n_{\mathrm{s}})$, and evolution parameters,
$\bm{\Theta}_{\mathrm{e}}$, using a two-stage process.

The first stage focuses on modelling the overall change in the broad-band shape of the non-linear power spectrum for a fixed
reference set of evolution parameters, $\bm{\Theta}_{\mathrm{e0}}$.
We define the boost factor, $B(k)$, as the ratio of the non-linear power
spectrum to its de-wiggled linear counterpart, $P_{\mathrm{DW}}(k)$, as
\begin{equation}
B(k | \bm{\Theta}_{\mathrm{s}}, \sigma_{12}) = \frac{P(k | \bm{\Theta}_{\mathrm{s}}, \bm{\Theta}_{\mathrm{e0}}, \sigma_{12})}{P_{\mathrm{DW}}(k | \bm{\Theta}_{\mathrm{s}}, \bm{\Theta}_{\mathrm{e0}}, \sigma_{12})}.
\label{eq:def_Bk}
\end{equation}
Here, $P_{\mathrm{DW}}(k)$ is a version of the linear power spectrum where the BAO signal is damped following the predictions of perturbation theory. It is constructed as
\begin{equation}
P_{\mathrm{DW}}(k) = P_{\mathrm{L}}(k) G(k) + P_{\mathrm{NW}}(k) [1 - G(k)].
\label{eq:def_Pdw}
\end{equation}
Here, $P_{\mathrm{NW}}(k)$ is a smoothed version of the linear power spectrum
that retains its broadband shape but has the BAO signal removed, computed using the 
discrete sine transform algorithm proposed in \citet{Hamann2010}. The function $G(k)$ is a Gaussian damping kernel,
\begin{equation}
G(k) = \exp\left(-\frac{1}{2} k^2 \sigma_v^2\right),
\label{eq:def_Gk}
\end{equation}
which depends on the linear velocity dispersion, $\sigma_v^2$, calculated from the linear power spectrum as
\begin{equation}
\sigma_v^2 = \frac{1}{6\pi^2} \int_0^\infty P_{\mathrm{L}}(k) \,{\rm d}k.
\label{eq:def_sigmav}
\end{equation}
This formulation smoothly transitions from the full linear power spectrum at large scales to the non-wiggled component at small scales.

We build an emulator of the boost factor, $\mathcal{E}_B(k)$, solely as a function of the shape parameters $\bm{\Theta}_{\mathrm{s}}$ and the clustering amplitude $\sigma_{12}$. Specifically, for each training cosmology defined by
$(\bm{\Theta}_{\mathrm{s}}, \sigma_{12})$, we run an N-body simulation with
evolution parameters $\bm{\Theta}_{\mathrm{e0}}$ until the redshift $z$ at which
$\sigma_{12}(z, \bm{\Theta}_{\mathrm{s}}, \bm{\Theta}_{\mathrm{e0}})$ matches
the target $\sigma_{12}$ of the node. In practice, rather than fixing the evolution parameter $\omega_{\mathrm{DE}}$
for $\bm{\Theta}_{\mathrm{e0}}$, we fix the value of $h$. Since $\omega_{\mathrm{b}}$ and $\omega_{\mathrm{c}}$ vary across the training set,
this implies that $\omega_{\mathrm{DE}}$ subtly varies to maintain the fixed $h$
for a flat universe (see Section~\ref{ssec:simulations}). 

\begin{figure}
	\includegraphics[width=0.95\columnwidth]{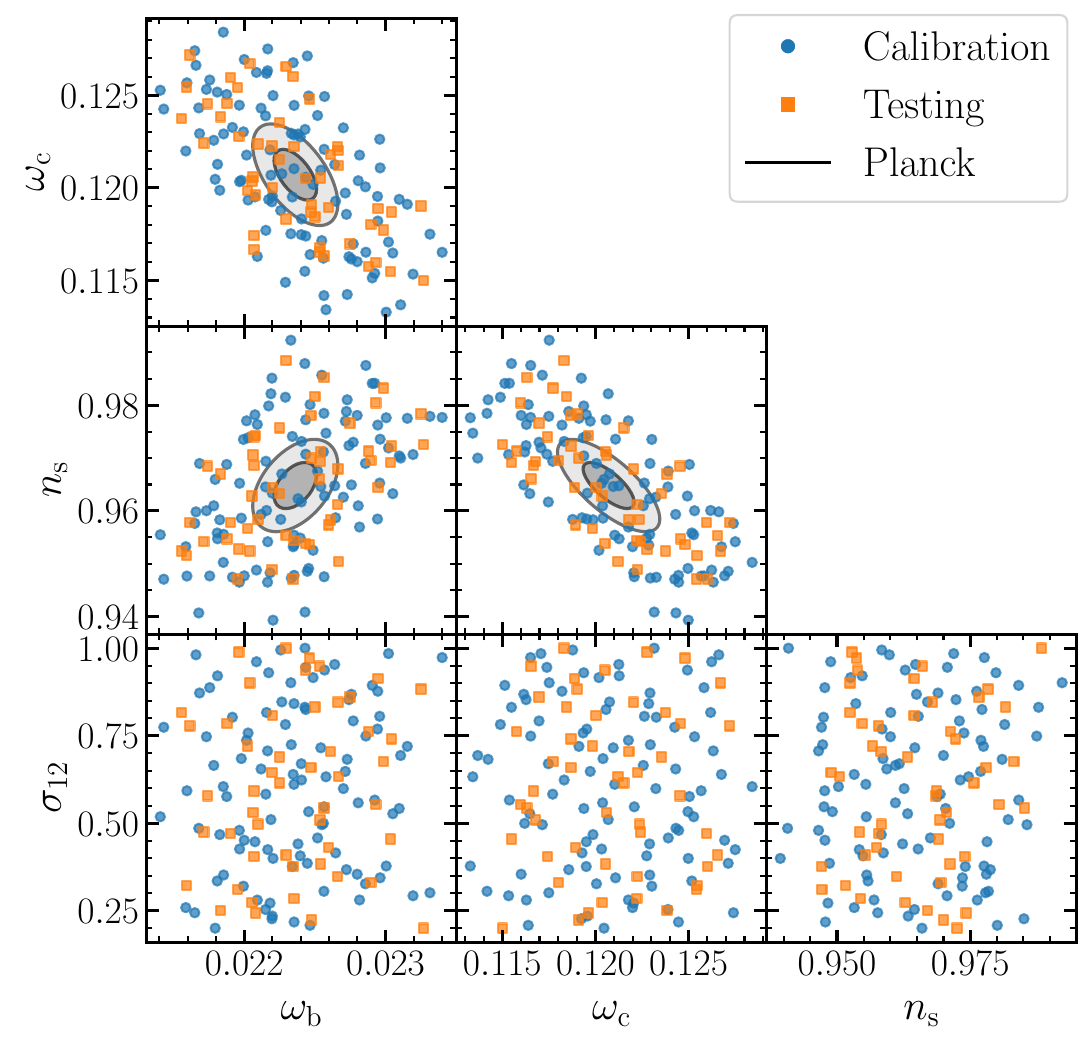}
    \caption{Distribution of cosmological parameters for the training (blue points) and testing (orange points) sets of simulations used for $\mathcal{E}_B(k)$. The panels show 2D projections of the parameter space $(\omega_{\mathrm{b}}, \omega_{\mathrm{c}}, n_{\mathrm{s}}, \sigma_{12})$. The grey ellipses represent the $1\sigma$ and $2\sigma$ confidence regions derived from \textit{Planck} 2018 data for the shape parameters $(\omega_{\mathrm{b}}, \omega_{\mathrm{c}}, n_{\mathrm{s}})$. Our sampling strategy uses the eigenvector directions of the \textit{Planck} covariance matrix to broadly cover the relevant parameter space. The sampling for $\sigma_{12}$ covers the range from $0.2$ to $1.0$.}
    \label{fig:aletheia_cosmologies}        
\end{figure}

The second stage in our prediction of $P(k)$ accounts for the small deviations that arise when the target cosmology has different evolution parameters
$\bm{\Theta}_{\mathrm{e}}$ than
the reference set $\bm{\Theta}_{\mathrm{e0}}$. We define the ratio
\begin{equation}
R(k | \bm{\Theta}_{\mathrm{s}}, \bm{\Theta}_{\mathrm{e}}, \sigma_{12}) = \frac{P(k | \bm{\Theta}_{\mathrm{s}}, \bm{\Theta}_{\mathrm{e}}, \sigma_{12})}{P(k | \bm{\Theta}_{\mathrm{s}}, \bm{\Theta}_{\mathrm{e0}}, \sigma_{12})},
\label{eq:def_Rk}
\end{equation}
where both power spectra are evaluated at the same
$\bm{\Theta}_{\mathrm{s}}$ and $\sigma_{12}$, but with differing evolution
parameters, leading to different integrated growth histories, $\tilde{x}$ and $\tilde{x}_0$. Following the linear approximation of equation~(\ref{eq:pk_taylor_xtilde}), we estimate this ratio as
\begin{equation}
R(k | \bm{\Theta}_{\mathrm{s}}, \bm{\Theta}_{\mathrm{e}}, \sigma_{12}) \approx 
1 + \left. \frac{\partial R(k)}{\partial\tilde{x}} \right|_{\tilde{x}_0} \left(\tilde{x} - \tilde{x}_0\right).
\label{eq:Rk_taylor}
\end{equation}
We build a second emulator, $\mathcal{E}_{\partial R/\partial\tilde{x}}$, to predict the derivative
term $\partial R(k)/\partial\tilde{x}$ as a function of $(\bm{\Theta}_{\mathrm{s}},
\sigma_{12})$.

The final prediction for the non-linear power spectrum for an arbitrary
cosmology $(\bm{\Theta}_{\mathrm{s}}, \bm{\Theta}_{\mathrm{e}})$ at a redshift
$z$, characterised by the clustering amplitude $\sigma_{12}$ and the integrated growth history parameter $\tilde{x}$, is given by
\begin{align}
\mathcal{E}_P(k|\bm{\Theta}_{\mathrm{s}}, \bm{\Theta}_{\mathrm{e}}, z) =\;
& P_{\mathrm{DW}}(k|\bm{\Theta}_{\mathrm{s}}, \bm{\Theta}_{\mathrm{e}}, z)
  \times \mathcal{E}_B(k|\bm{\Theta}_{\mathrm{s}}, \sigma_{12}) \notag \\
& \times \left[1 + \mathcal{E}_{\partial R/\partial\tilde{x}}(k|\bm{\Theta}_{\mathrm{s}}, \sigma_{12}) \left(\tilde{x} - \tilde{x}_0\right) \right].
\label{eq:pk_final_aletheia}
\end{align}
Here, $P_{\mathrm{DW}}(k|\bm{\Theta}_{\mathrm{s}}, \bm{\Theta}_{\mathrm{e}}, z)$ is the de-wiggled linear power spectrum of the target cosmology and $\tilde{x}_0$ is the value 
of the integrated growth history parameter for the reference evolution parameters $\bm{\Theta}_{\mathrm{e0}}$ evaluated at the redshift $z_0$ that corresponds to the same clustering amplitude, i.e., $\sigma_{12}(z_0, \bm{\Theta}_{\mathrm{s}}, \bm{\Theta}_{\mathrm{e0}}) = \sigma_{12}(z, \bm{\Theta}_{\mathrm{s}}, \bm{\Theta}_{\mathrm{e}})$.

By construction, this design ensures that the emulator is broadly applicable to
various evolution histories, including dynamic dark energy models, without
explicitly sampling their parameters or redshift during the primary training of
the emulators $\mathcal{E}_B$ and $\mathcal{E}_{dR/dx}$, which only depend on
$(\omega_{\mathrm{b}}, \omega_{\mathrm{c}}, n_{\mathrm{s}}, \sigma_{12}, \ln k)$.
The value of $\ln k$ is treated as an input parameter for both emulators, allowing
their evaluation at any desired $k$ value. The ranges for these parameters will be
detailed in Section~\ref{ssec:simulations}. 

\subsection{Cosmological simulations}
\label{ssec:simulations}

Two different sets of N-body simulations, which we collectively refer to as the AletheiaEmu suite, were performed to
train and test the {\tt Aletheia} emulators. All simulations use initial
conditions generated at $z_\mathrm{IC}=99$ with \textsc{2LPTic}
\citep{crocce_2lptic,Crocce2012_code2lptic} and evolved to the desired redshift
using the publicly available version of the code \textsc{Gadget-4}  \citep{Springel2021_Gadget4}. Both codes were
modified to incorporate various dark energy models at the level of the background expansion. The Plummer-equivalent
softening length was set appropriately for the resolution of each simulation
set.

\begin{figure*}
\includegraphics[width=0.98\textwidth]{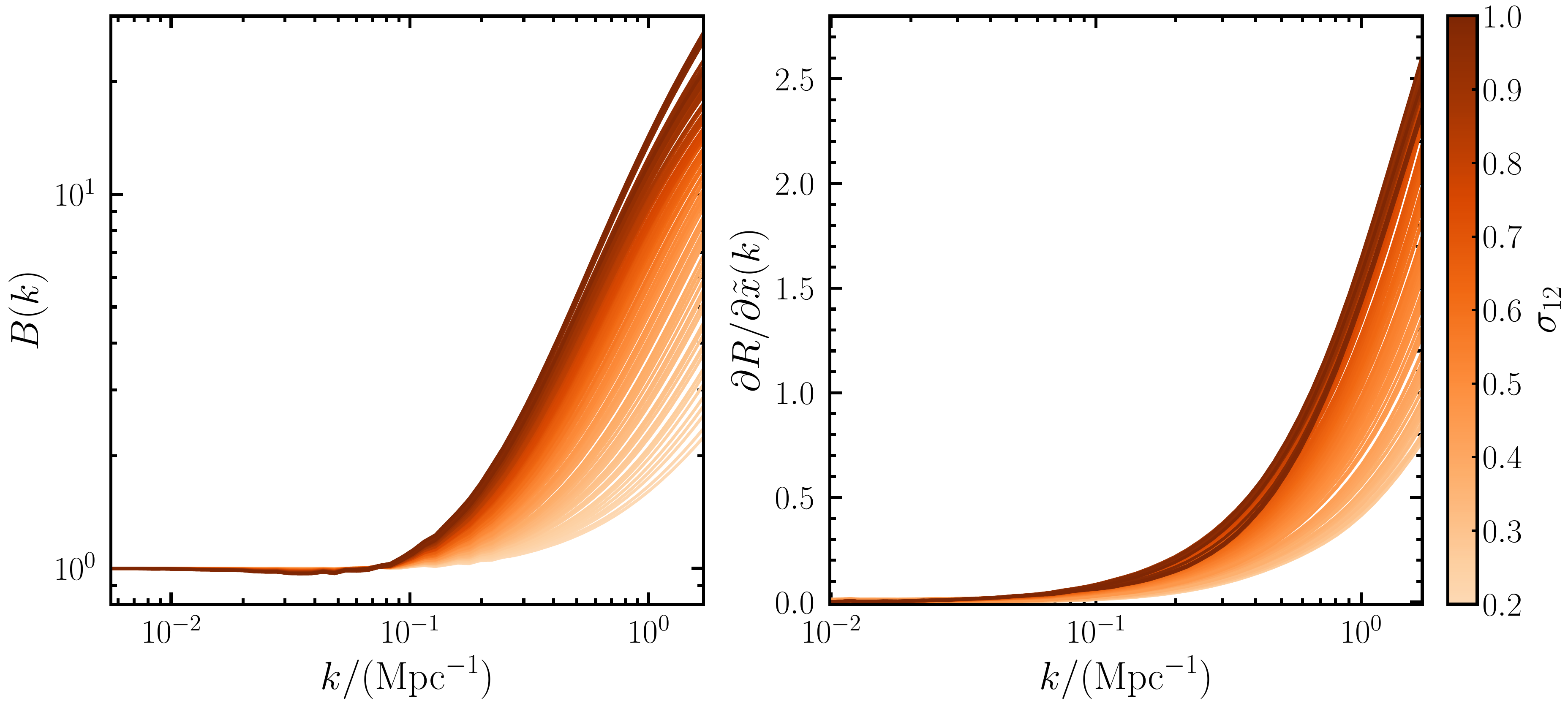}
    \caption{Raw training data for the Aletheia emulators. The left panel shows the measured boost factor, $B(k)$, for the 100 simulations to train $\mathcal{E}_B(k)$, colour-coded by the clustering amplitude $\sigma_{12}$ at which they were evaluated. This panel illustrates the strong, smooth dependence of the boost factor on $\sigma_{12}$. The right panel displays the measured derivative term $\partial R/\partial \tilde{x}$ for the same cosmologies, determined from the derivative simulations described in Section~\ref{sssec:sims_dRdx}. The smooth dependency of both quantities on the cosmological parameters and $k$ makes them well-suited for Gaussian Process emulation.}
    \label{fig:raw_training}        
\end{figure*}

\subsubsection{Simulations for $\mathcal{E}_B(k)$}
\label{sssec:sims_B}

To train the emulator $\mathcal{E}_B(k)$, we generated a set of 100 cosmological
models using a four-dimensional maximin Latin hypercube (LH) design \citep{mckay1979_LH,stein1987_MLH}. The sampling of the shape parameters $(\omega_{\mathrm{b}}, \omega_{\mathrm{c}}, n_{\mathrm{s}})$ for these models was designed to efficiently cover the physically relevant parameter space, based on current cosmological constraints. Rather than directly sampling these parameters, we used the covariance matrix of constraints derived from \textit{Planck} data \citep{Planck2018}. These constraints are largely robust to assumptions about dark energy evolution, making them suitable for defining the shape parameter space. 

We computed the eigenvalues and eigenvectors of this covariance matrix, which define the principal axes and their corresponding variances in the parameter space. The first three components of our LH sampling were identified with these eigenvector directions. For each eigenvector direction, the sampling range was set to $\pm5$ times the square root of its corresponding eigenvalue, ensuring a broad coverage that far exceeds the constraints from \textit{Planck}. The fourth parameter in our LH design was associated with $\sigma_{12}$ and sampled from 0.2 to 1.0, covering a wide range of relevant clustering amplitudes. The distribution of these 100 training cosmologies, along with their relationship to the \textit{Planck} 2018 constraints, is illustrated in Fig.~\ref{fig:aletheia_cosmologies}.

For each of these 100 nodes, the primordial scalar amplitude was kept fixed to
$A_{\mathrm{s}} = 2.1\times 10^{-9}$, and the dimensionless Hubble parameter was
fixed to $h =0.673$. As $\omega_{\mathrm{b}}$ and $\omega_{\mathrm{c}}$ vary across the training set, fixing $h$ implies that the dark energy density parameter, $\omega_{\mathrm{DE}}$, varies slightly across these simulations to maintain a flat universe. This approach defines our reference evolution parameters $\bm{\Theta}_{\mathrm{e0}}$ for each node. Each simulation was run until the specific redshift $z_i$ at which $\sigma_{12}(z_i)$ for that cosmology matched
the target $\sigma_{12,i}$ of the corresponding LH node. Thus, each of these
primary simulations yields a single snapshot used for training
$\mathcal{E}_B(k)$.  

For each LH node, we ran two N-body simulations employing the
``fixed-paired'' technique to suppress cosmic variance \citep{Angulo2016}. These
simulations follow the evolution of $2048^3$ dark matter particles in a periodic
box of side length $L_{\mathrm{box}} = 1500 \, \mathrm{Mpc}$. All simulation pairs use the same initial random phases to
minimise stochastic differences among the nodes. 

The non-linear matter power spectra of these simulations were measured in Mpc units using a Cloud-In-Cell \citep[CIC, ][]{Hockney1988} mass assignment scheme on a grid of 2048 points per side.  We corrected for the effect of aliasing with the interlacing method of \cite{Sefusatti2016}. This provides reliable measurements of $P(k)$ up to the Nyquist frequency, $k_{\rm Ny} =4.3\,{\rm Mpc}^{-1}$. The power spectra from the paired simulations were averaged to obtain the final estimate of the non-linear $P(k)$ for each node. Finally, these estimates were used to compute the boost factor, $B(k)$, for each node by dividing them by the corresponding de-wiggled power spectrum $P_{\rm DW}(k)$, as defined in equation~(\ref{eq:def_Bk}). We focus our emulation on the range $0.006\,{\rm Mpc}^{-1}<  k < 2\,{\rm Mpc}^{-1}$. The upper bound is chosen to mitigate the impact of resolution effects, for which we apply a correction to the final emulator results (see Section \ref{sec:resolution}). The left panel of Fig.~\ref{fig:raw_training} shows the measured $B(k)$ for all 100 training cosmologies, colour-coded by their corresponding values of $\sigma_{12}$, illustrating the primary dependence captured by $\mathcal{E}_B(k)$.

An additional set of 50 cosmological models was generated for testing the performance of the emulator. These testing cosmologies were sampled using a different maximin LH, but followed the same eigenvector-based parameter space definition, fixed values of $A_\mathrm{s}$ and $h$, target $\sigma_{12}$ selection, and paired simulation specifications with $2048^3$ particles. The distribution of these 50 testing cosmologies is also shown in Fig.~\ref{fig:aletheia_cosmologies}.

\subsubsection{Simulations for $\mathcal{E}_{\partial R/\partial \tilde{x}}(k)$}
\label{sssec:sims_dRdx}

To train our second emulator $\mathcal{E}_{\partial R/\partial\tilde{x}}(k)$, we used the same set of 100 base cosmological models described in Section~\ref{sssec:sims_B}. For each node, we ran five additional simulations for cosmological models characterised by the same shape parameters of the main node, but exploring variations in the evolution parameters. Their purpose is to measure the response of the non-linear power spectrum to variations in the integrated growth history parameter $\tilde{x}$. These additional cosmologies include:
\begin{enumerate}
\item One simulation with the same base evolution parameters ($\bm{\Theta}_{\mathrm{e0}}$), serving as the reference $P_0(k)$.
\item Two simulations where the dark energy density is varied relative to the base value as $\omega_{\mathrm{DE}}=0.925\,\omega_{\mathrm{DE},0}$ and $\omega_{\mathrm{DE}}=1.125\,\omega_{\mathrm{DE},0}$.
\item Two simulations where the physical curvature density is set to $\omega_{K}=-0.04$ and $\omega_{K}=+0.02$.
\end{enumerate}
By inducing these targeted variations in the evolution parameters, we generate models with different integrated growth histories ($\tilde{x} \ne \tilde{x}_0$) while keeping the clustering amplitude $\sigma_{12}$ fixed.

All five simulations for each node were run with $1500^3$ particles in a box of size $L_{\mathrm{box}} = 1500 \, \mathrm{Mpc}$ and used the same initial random phases. The power spectra of each simulation was measured using the same method described in Section~\ref{sssec:sims_B}.

These measurements were used to define the ratios $R_j(k) = P_j(k)/P_0(k)$, where $P_0(k)$ is the power spectrum of the reference simulation ($\bm{\Theta}_{\mathrm{e0}}$). Assuming the linear approximation established in equation~(\ref{eq:pk_taylor_xtilde}), the derivative $\partial R/\partial \tilde{x}(k)$ is then obtained by performing a linear fit to the measured $R_j(k)$ versus $\Delta\tilde{x}=(\tilde{x}_j - \tilde{x}_0)$, constrained to $R(k)=1$ when $\Delta\tilde{x} = 0$.

Since the derivative term $\partial R/\partial \tilde{x}$ is a small relative correction factor, it does not require the same demanding fidelity as the absolute $P(k)$ measurements for $\mathcal{E}_B(k)$. Furthermore, we verified that the measured power spectrum ratios, $R_j(k)$, are insensitive to the resolution of the simulations. 
This allowed us to use a lower resolution for this second set of simulations than the ones used for the training of the primary emulator $\mathcal{E}_B(k)$, significantly reducing their computational cost. 

The set of 100 derivative measurements was used to train the emulator $\mathcal{E}_{\partial R/\partial \tilde{x}}(k)$. As illustrated in the right panel of Fig.~\ref{fig:raw_training}, this derivative term is a smooth function of $k$ and the input parameters. The same procedure is applied to the 50 test cosmologies described in Section~\ref{sssec:sims_B} by running the corresponding derivative simulations, providing an independent set of measurements for validation.

\begin{figure}
    \includegraphics[width=0.98\columnwidth]{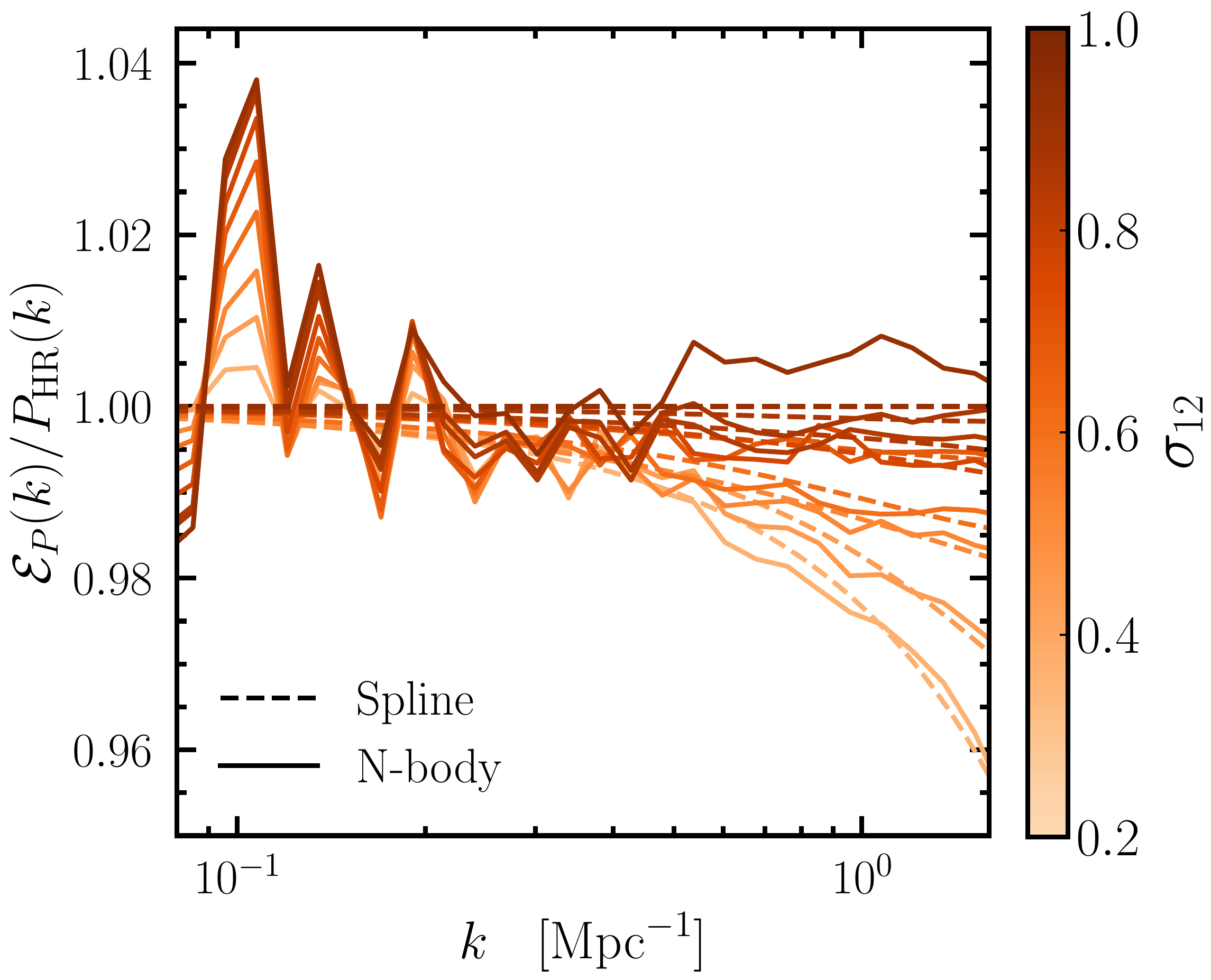}
    \caption{Correction of the \texttt{Aletheia} emulator for resolution effects. The solid lines show the measured ratio of the uncorrected emulator prediction, $\mathcal{E}_P(k)$, to the high-resolution power spectrum, $P_{\mathrm{HR}}(k)$, derived from the AletheiaMass simulations. The lines are colour-coded by the value of $\sigma_{12}$. The fluctuations are caused by cosmic variance due to the smaller box size of the high-resolution runs. The dashed lines represent the corresponding two-dimensional smoothing spline interpolation, $\mathcal{C}(k, \sigma_{12})$, which is used to apply the correction to the final emulator prediction.}
    \label{fig:resolution_correction}
\end{figure}

\subsection{Gaussian process emulation}
\label{ssec:gp_emulation}

Both $\mathcal{E}_B(k)$ and $\mathcal{E}_{\partial R/\partial\tilde{x}}(k)$ are constructed using
Gaussian Process (GP) regression, a powerful non-parametric Bayesian method for
interpolating complex functions from a finite set of training data
\citep{Rasmussen2006_GPML}. A GP defines a probability distribution over
functions. Given a set of training inputs $\mathbf{X} = \{\mathbf{x}_1, ...,
\mathbf{x}_N\}$ and corresponding outputs $\mathbf{y} = \{y_1, ..., y_N\}$, a GP
model can predict the output $y$ at a new input point $\mathbf{x}$
by providing a mean prediction and an estimate of the prediction uncertainty.
The key component of a GP is the covariance function, or kernel,
$K(\mathbf{x}_i, \mathbf{x}_j)$, which defines the similarity between output
values based on their input points. 

For both our emulators, $\mathcal{E}_B(k)$ and $\mathcal{E}_{\partial R/\partial\tilde{x}}(k)$, we employ
a Mat\'ern kernel with a smoothness parameter $\nu=3/2$.  This kernel takes the form 
\begin{equation}
K_{\nu=3/2}(r) = \sigma_f^2 \left(1 + \sqrt{3}r\right) \exp\left(-\sqrt{3}r\right),
\label{eq:matern32}
\end{equation}
where $\sigma_f^2$ is the signal variance, $r$ is the anisotropic Mahalanobis distance between input points $\mathbf{x}$ and $\mathbf{x}'$ in $D$ dimensions, defined as
\begin{equation}
    r = \sqrt{\sum_{i=1}^{D} \left(\frac{x_i - x'_i}{\ell_i}\right)^2},
    \label{eq:r_dist}
\end{equation}
with $\ell_i$ giving the characteristic length-scales for each input dimension. The hyperparameters  $\sigma_f^2$ and $\ell_i$ were optimised by maximising the
marginal likelihood of the training data.

The input parameters for both emulators are ($\omega_{\mathrm{b}}$, $\omega_{\mathrm{c}}$, $n_{\mathrm{s}}$, $\sigma_{12}$, $\ln(k)$). As mentioned before, 
$\ln(k)$ is treated as an explicit input parameter, allowing the emulators to be queried at any arbitrary $k$ value within the trained range, rather than being trained independently at a fixed grid of $k$-bins, offering more flexibility.

The training of the GPs for {\tt Aletheia} was performed using the
\texttt{scikit-learn} Python library \citep{Pedregosa2011_scikitlearn}. The
training process involves finding the optimal kernel hyperparameters, and the
testing process (on the 50 separate cosmologies) evaluates the predictive
accuracy of the emulators and quantifies their uncertainties.

\subsection{Correction for resolution effects}
\label{sec:resolution}

The prediction of the non-linear power spectrum from our combined emulator, $\mathcal{E}_P(k)$, is inherently limited by the resolution of the AletheiaEmu simulations of the primary training, which used $2048^3$ particles in a $1500\,{\rm Mpc}$ box. At high $k$, these predictions exhibit a suppression of power due to the finite particle resolution. To extend the accuracy of \texttt{Aletheia} more deeply into the non-linear regime, we implemented a correction factor based on higher-resolution simulations.

\begin{figure*}
    \includegraphics[width=0.95\textwidth]{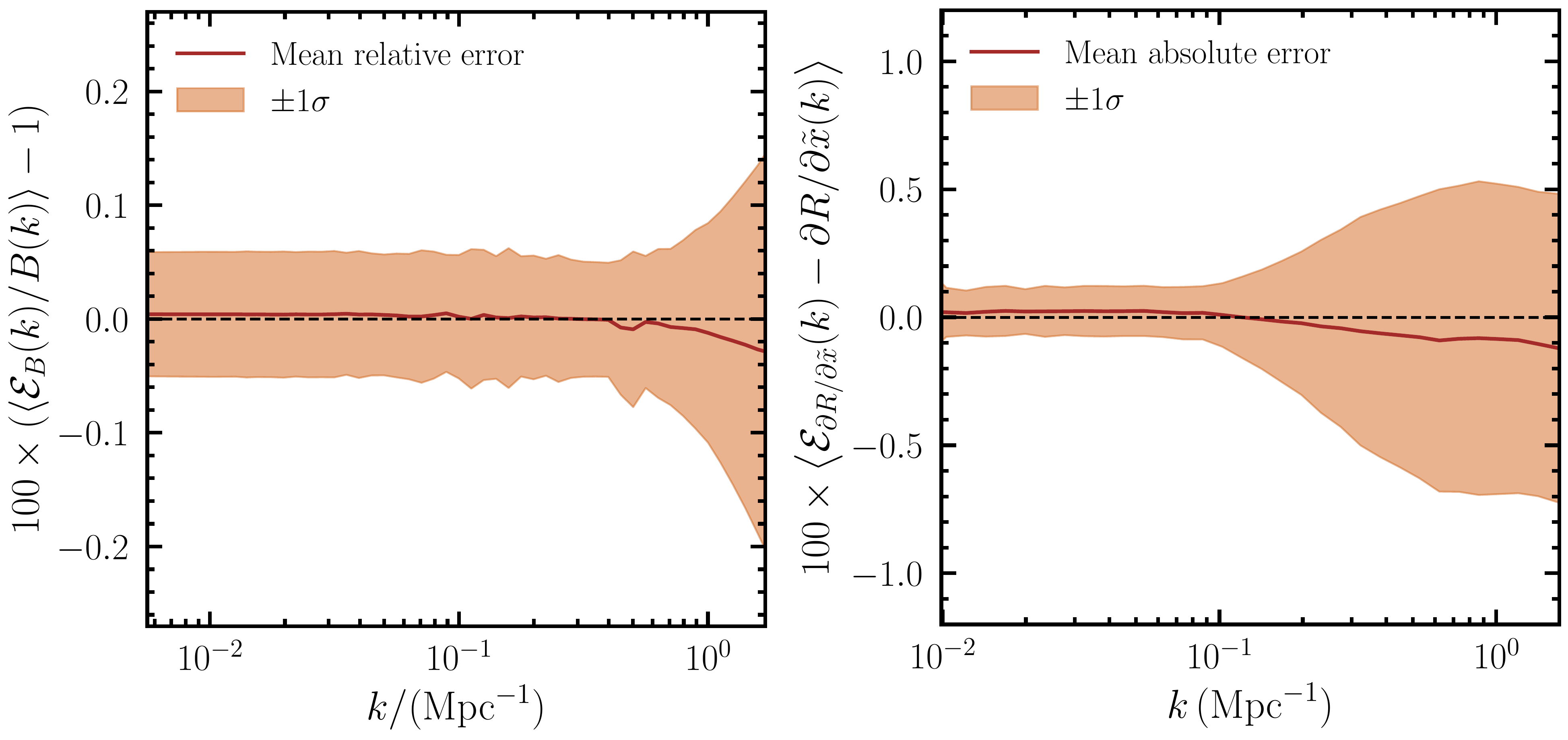}
    \caption{Performance of the individual emulator components of \texttt{Aletheia}. The left panel
    shows the relative error of the $\mathcal{E}_B(k)$ emulator,
    $(P_{\mathrm{emu}}/P_{\mathrm{sim}} - 1)$, with the shaded region indicating
    the $1\sigma$ variance across the test set. The right panel displays the
    absolute error of the $\mathcal{E}_{\mathrm{d}R/\mathrm{d}\tilde{x}}(k)$
    emulator. Both emulators demonstrate high accuracy across the relevant
    wavenumber range, with the error of $\mathcal{E}_B(k)$ being consistently
    below 0.2\% and the absolute error of
    $\mathcal{E}_{\mathrm{d}R/\mathrm{d}\tilde{x}}(k)$ being very small.}
    \label{fig:performance_individual_emus}  
\end{figure*}

This correction is defined using the AletheiaMass suite of N-body simulations, described in detail in Fiorilli et al. (in prep.). These are a set of pair-fixed runs for a cosmology close to the best-fitting $\Lambda$CDM model of \citet{Planck2018} that sample the growth of structure over a range of $\sigma_{12}$ matching that covered by our emulator ($0.2$ to $1.0$). 
We obtained high-resolution power spectra, $P_{\mathrm{HR}}(k)$, by extracting data from the AletheiaMass runs with $2048^3$ particles in a  $700\,{\rm Mpc}$ box. This configuration results in significantly higher mass resolution than the AletheiaEmu training suite.

We define the resolution correction factor, $\mathcal{C}(k, \sigma_{12})$, as the ratio of the uncorrected emulator prediction to the higher-resolution power spectrum,
\begin{equation}
\mathcal{C}(k, \sigma_{12}) = \frac{\mathcal{E}_P(k| \bm{\Theta}_{\mathrm{fid}}, \sigma_{12})}{P_{\mathrm{HR}}(k| \bm{\Theta}_{\mathrm{fid}}, \sigma_{12})}.
\label{eq:resolution_correction}
\end{equation}
We assume that this correction is independent of the shape parameters and depends only on the wavenumber $k$ and the clustering amplitude $\sigma_{12}$. 
Fig.~\ref{fig:resolution_correction} shows the measured correction ratios $\mathcal{C}(k, \sigma_{12})$ as solid lines, colour-coded by $\sigma_{12}$. We model this ratio using a two-dimensional smoothing spline over the $(k, \sigma_{12})$ plane. The correction is assumed to be unity for high clustering amplitudes, specifically $\sigma_{12} > 0.83$. The dashed lines in Fig.~\ref{fig:resolution_correction} illustrate the resulting smoothed spline interpolation, $\mathcal{C}_{\mathrm{s}}(k, \sigma_{12})$. The final corrected prediction of the emulator, $\mathcal{E}_{\mathrm{final}}(k)$, is then given by:
\begin{equation}
\mathcal{E}_{\mathrm{final}}(k) = \frac{\mathcal{E}_P(k)}{ \mathcal{C}_{\mathrm{s}}(k, \sigma_{12})}.
\label{eq:final_emulator_prediction}
\end{equation}
This is the final recipe implemented in the public \texttt{Aletheia} Python package (see Appendix~\ref{sec:appendix_code}).

\section{Validation of the emulators}
\label{sec:validation}

An essential step in the development of any emulator is a rigorous validation of its performance against independent data not used during the training phase. In this section, we present a comprehensive assessment of \texttt{Aletheia}, starting with the individual components of our two-stage emulation strategy and then moving on to the validation of the full, combined emulator.

\subsection{Validation of individual components}
\label{ssec:validation_components}

We trained the emulators $\mathcal{E}_B(k)$ and $\mathcal{E}_{\partial R/\partial\tilde{x}}(k)$ on the sets of simulations described in Section~\ref{ssec:simulations}. To obtain an unbiased assessment of their accuracy, we validated their performance using a dedicated set of 50 testing cosmologies, which were entirely excluded from the training process.

Figure~\ref{fig:performance_individual_emus} illustrates the performance of both emulator components. The left panel shows the relative error of the emulator $\mathcal{E}_B(k)$ with respect to the boost factors $B(k)$ measured in the test simulations. The solid line indicates the mean relative error across all testing cosmologies, and the shaded region shows the corresponding $1\sigma$ variance. The mean relative error is remarkably small, remaining consistently below $0.03\%$ across the full range of scales considered. The variance of the predictions is stable at the level of $0.05\%$ for $k \leq 0.8\,{\rm Mpc}^{-1}$, and gradually increases for larger wavemodes while staying below $0.2\%$, demonstrating excellent accuracy.

The right panel of Fig.~\ref{fig:performance_individual_emus} presents the absolute error of the $\mathcal{E}_{\partial R/\partial\tilde{x}}(k)$ emulator. We chose to show the absolute error for this component because the derivative $\partial R/\partial\tilde{x}(k)$ approaches zero at low $k$, making relative errors ill-suited for a clear visual representation. The solid line shows the mean absolute error across the testing cosmologies, and the shaded region indicates the $1\sigma$ variance. The mean absolute error for $\mathcal{E}_{\partial R/\partial\tilde{x}}(k)$ is generally below $0.007$ at high wavenumbers. This absolute error typically corresponds to less than $0.3\%$ of the maximum value of the derivative on those scales.

Both emulators demonstrate good performance, well within the requirements for precision cosmology. It is worth noting that the emulator for the derivative term, $\mathcal{E}_{\partial R/\partial\tilde{x}}(k)$, does not require the same level of stringent accuracy as $\mathcal{E}_B(k)$. This is because the predicted derivative is multiplied by the term $(\tilde{x} - \tilde{x}_0)$ in equation~(\ref{eq:pk_final_aletheia}), which is typically of the order of $10^{-2}$. Therefore, a small percentage error in $\partial R/\partial\tilde{x}$ translates into an even smaller fractional error in the final power spectrum prediction. The high performance of these individual components provides strong confidence in the overall accuracy of the \texttt{Aletheia} emulator.

\subsection{Validation of the full \texttt{Aletheia} emulator}
\label{ssec:validation_full_emu}

Having validated the individual components of \texttt{Aletheia}, we now present an assessment of its performance in predicting the full non-linear matter power spectrum $P(k)$ for diverse cosmological scenarios. We compare \texttt{Aletheia}'s predictions against a set of independent N-body simulations, which we collectively refer to as the AletheiaDE test suite. We benchmark its accuracy against the results of \texttt{EuclidEmulator2} \citep{EuclidEmulator2}. We choose \texttt{EuclidEmulator2} as a representative of the state of the art in the field due to its wide adoption, well-documented performance,  extended parameter space, and broad redshift range.

\begin{table*}
    \caption{Cosmological parameters for the AletheiaDE simulation set. This
    table lists the physical baryon ($\omega_{\mathrm{b}}$), and cold dark
    matter ($\omega_{\mathrm{c}}$) densities, the primordial scalar spectral
    index ($n_{\mathrm{s}}$), the physical dark energy density
    ($\omega_{\mathrm{DE}}$) and its equation of state parameters ($w_0, w_a$),
    the dimensionless Hubble parameter ($h$), and the values of $\sigma_{12}$
    and $\sigma_{8/h}$ at redshift $z=0$. The first ten rows correspond to
    cosmologies within the parameter ranges of {\tt EuclidEmulator2}. The
    last row represents the best-fit dynamic dark energy cosmology to DESI DR1
    BAO data, combined with CMB and SN measurements \citep{Adame2025_DESIBAO1}.}
    \label{tab:test_cosmologies}
    \centering
    \begin{tabular}{cccccccccc}
        \hline
        ID & $\omega_{\mathrm{b}}$ & $\omega_{\mathrm{c}}$ & $n_{\mathrm{s}}$ & $\omega_{\mathrm{DE}}$ 
        & $w_0$ & $w_a$ & $h$ & $\sigma_{12}(z=0)$  & $\sigma_{8/h}(z=0)$ \\
        \hline
         1 & 0.02213 & 0.11671  & 0.9837 & 0.2865 & -1.1118 &  0.4722 & 0.6522 & 0.8144  & 0.8021 \\
         2 & 0.02219 & 0.12459  & 0.9488 & 0.2308 & -0.8340 & -0.2718 & 0.6145 & 0.9291  & 0.8774 \\
         3 & 0.02136 & 0.12821  & 0.9474 & 0.3136 & -0.9505 &  0.3855 & 0.6806 & 0.8296  & 0.8416 \\
         4 & 0.02207 & 0.11665  & 0.9806 & 0.3433 & -0.7545 & -0.3737 & 0.6943 & 0.7741 & 0.7961 \\
         5 & 0.02277 & 0.12048  & 0.9628 & 0.2854 & -1.0852 &  0.3930 & 0.6547 & 0.7513 & 0.7419 \\
         6 & 0.02236 & 0.12269  & 0.9461 & 0.3461 & -0.9064 &  0.2483 & 0.7008 & 0.7804 & 0.8076 \\
         7 & 0.02249 & 0.11512  & 0.9699 & 0.2357 & -1.0547 &  0.1569 & 0.6110 & 0.7496 & 0.7609 \\
         8 & 0.02309 & 0.11680  & 0.9645 & 0.2571 & -1.2528 & -0.6151 & 0.6301 & 0.8087 & 0.7778 \\
         9 & 0.02167 & 0.12578  & 0.9570 & 0.3833 & -0.9794 &  0.3158 & 0.7285 & 0.7464 & 0.7937 \\
        10 & 0.02229 & 0.11652  & 0.9762 & 0.3100 & -1.0000 &  0.0000 & 0.6700 & 0.8078 & 0.8106 \\
        \hline
        DESI Y1 Best-fit & 0.02236 & 0.12021 & 0.9648 & 0.3000 & -0.6500 & -1.2700 & 0.6653 & 0.8230 & 0.8218 \\ 
        \hline
    \end{tabular}
\end{table*}

For this validation, we selected ten distinct dynamic dark energy cosmological models, including one $\Lambda$CDM model ($w_0 = -1, w_a=0$). These models were chosen such that their parameters fall within the parameter space covered by \texttt{EuclidEmulator2}, and their shape parameters ($\omega_{\mathrm{b}}, \omega_{\mathrm{c}}, n_{\mathrm{s}}$) are within the prior ranges of \texttt{Aletheia}  described in Section~\ref{ssec:simulations}. The full list of cosmological parameters defining these cosmologies is summarised in Table~\ref{tab:test_cosmologies}.

For each of these cosmologies, we ran two paired and fixed N-body simulations, using initial phases distinct from those employed for the training simulations. These simulations used $2048^3$ particles in a periodic box of size $L_{\rm box}=1500\,{\rm Mpc}$. We obtained snapshots at six different redshifts: $z = 3, 2, 1, 0.6, 0.3,$ and $0$, spanning the full redshift range covered by \texttt{EuclidEmulator2}. For each cosmology and redshift, we estimated the matter power spectrum $P(k)$ following the same methodology as described in Section~\ref{sssec:sims_B} and then obtained the corresponding predictions from both \texttt{Aletheia} and \texttt{EuclidEmulator2}.

\begin{figure}
\includegraphics[width=0.95\columnwidth]{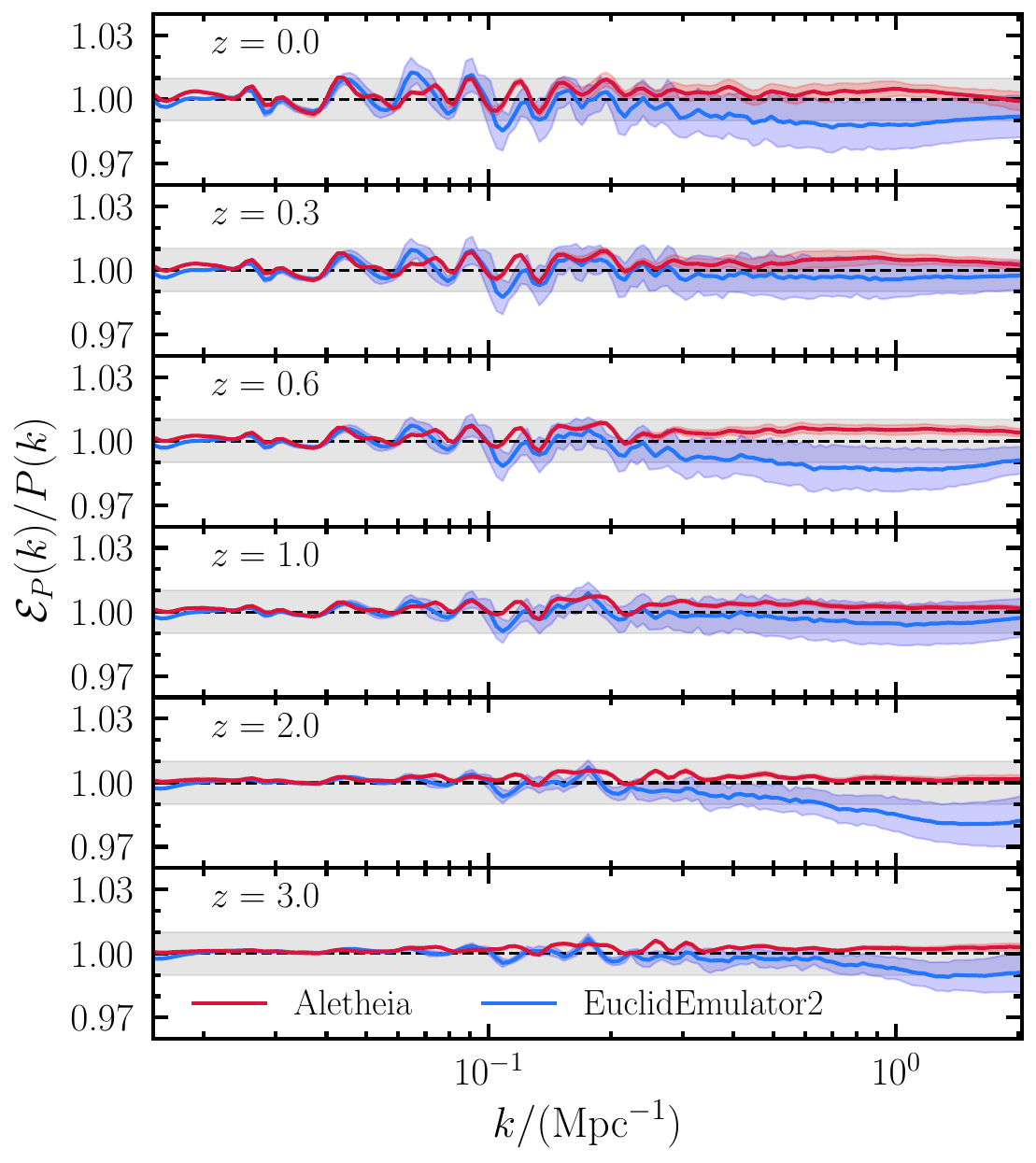}
    \caption{
    Comparison of the performance of the full \texttt{Aletheia} emulator against N-body simulations and \texttt{EuclidEmulator2}. Each panel shows the mean and variance of the ratio $\mathcal{E}_P(k)/P(k)$ across ten test cosmologies for various redshifts computed using \texttt{Aletheia} (red) and \texttt{EuclidEmulator2} (blue). Both emulators show excellent agreement with the simulations, with deviations largely driven by cosmic variance. \texttt{Aletheia} exhibits significantly lower variance than \texttt{EuclidEmulator2}, illustrating the higher accuracy enabled by the evolution mapping framework.}
\label{fig:performance_euclid_comparison} 
\end{figure}
 
Figure~\ref{fig:performance_euclid_comparison} presents the comparison of the performance of both emulators. Each panel corresponds to a different redshift and shows the mean and variance of the ratio $\mathcal{E}_P(k)/P(k)$ for both \texttt{Aletheia} (red lines) and \texttt{EuclidEmulator2} (blue lines) across the ten test cosmologies. 
Note that the AletheiaDE simulations used for this test were run at the same resolution as the primary emulator training suite ($2048^3$ particles in $1500\,{\rm Mpc}$ boxes), therefore, no resolution correction was applied to the \texttt{Aletheia} predictions shown here. 
The results demonstrate good agreement between both emulators and the N-body simulation results, with overall percent-level differences. Both emulators exhibit similar deviations from the simulation measurements, which are primarily due to cosmic variance associated with the particular initial phases of the simulations. This indicates that both emulators are capturing the underlying physics robustly.

Although the mean ratio curves for both emulators are close to unity, a notable difference can be appreciated in the variance of the predictions. While the variance of the predictions of \texttt{EuclidEmulator2} is of the order of $1\%$, \texttt{Aletheia} shows a variance of approximately $0.2\%$. Although this result is based on a relatively small set of test simulations, this significantly lower variance confirms the power of the evolution mapping approach: the compressed parameter space of the evolution mapping framework results in more accurate and stable predictions compared to those obtained with the standard approach of directly emulating the dependency on all evolution parameters and redshift.

A further advantage of the evolution mapping emulation strategy is its potential for broad generalisation to much wider ranges of redshifts and cosmological parameters, while still maintaining high accuracy. To illustrate this, we performed an additional test using a cosmology explicitly outside the typical parameter space of state-of-the-art emulators. We ran an additional pair of fixed-paired simulations corresponding to the best-fitting dynamical dark energy cosmology obtained from a joint fit to DESI DR1 BAO data, CMB measurements, and SN data \citep{Adame2025_DESIBAO1}. The parameters for this specific cosmology are listed in the last row of Table~\ref{tab:test_cosmologies}. This cosmology is outside the limits of most existing emulators (for example, its $w_a$ value is almost twice the prior range allowed by \texttt{EuclidEmulator2}). Our \texttt{Aletheia} emulator, however, can reproduce accurately the results of this simulation, illustrating the greater generality of the evolution mapping approach.

\begin{figure} 
    \includegraphics[width=0.95\columnwidth]{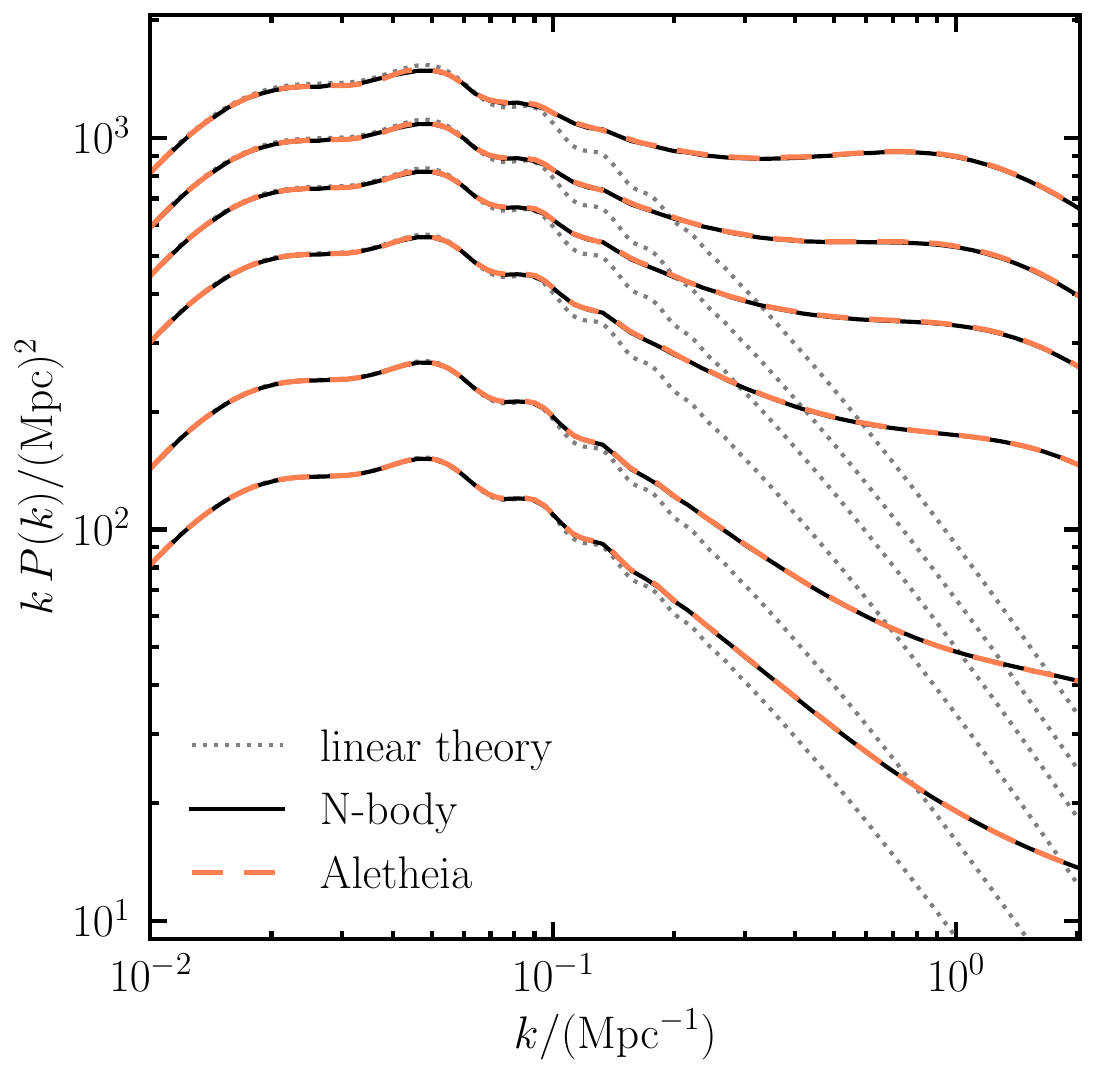}
    \caption{
    Non-linear matter power spectrum for the best-fitting dynamic dark energy cosmology of \citet{Adame2025_DESIBAO1}. N-body simulation results (black solid lines) at six redshifts are compared against the predictions of \texttt{Aletheia} (red dashed lines). The linear-theory predictions (grey dotted lines) illustrate the impact of non-linear evolution. 
    The evolution mapping design allows \texttt{Aletheia} to provide reliable predictions for cosmologies outside the parameter space of conventional emulators.
    } 
    \label{fig:aletheia_desi_comparison_pk}
\end{figure}

\begin{figure}  \includegraphics[width=0.95\columnwidth]{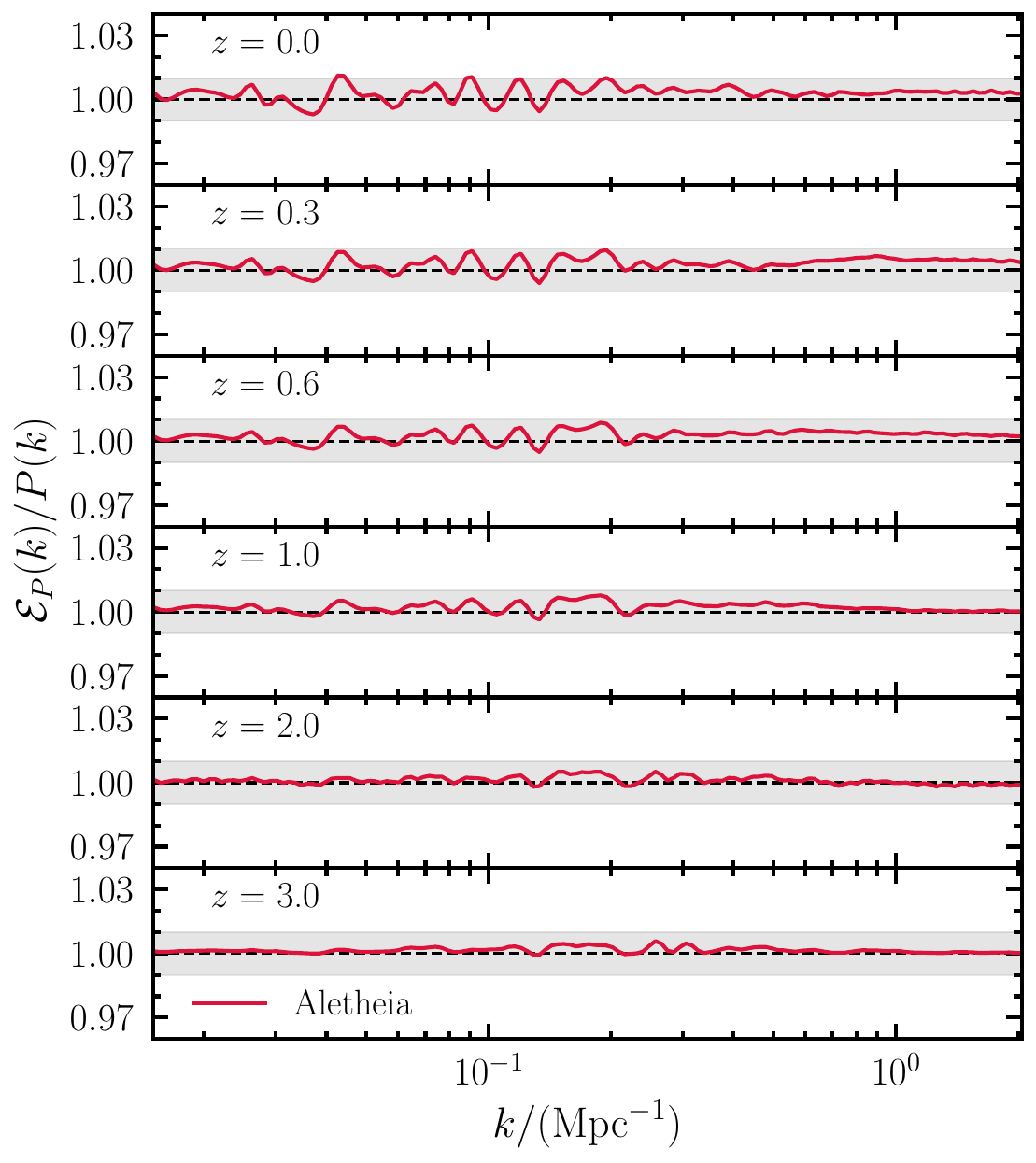}
    \caption{
    Ratio of \texttt{Aletheia}'s prediction to N-body simulation results for the best-fit dynamic dark energy cosmology of \citet{Adame2025_DESIBAO1}. The red solid lines in each panel show $\mathcal{E}_P(k) / P(k)$ for six different redshifts. The grey shaded regions indicates a $\pm 1\%$ agreement band.
    The accuracy achieved demonstrates the validity of the evolution mapping design over a significantly wider cosmological parameter space than conventional emulators.
    }  
    \label{fig:aletheia_desi_comparison_ratio} 
\end{figure}

Figure~\ref{fig:aletheia_desi_comparison_pk} shows the comparison of the measured $P(k)$ from these DESI-motivated simulations (black solid lines) with \texttt{Aletheia}'s predictions (dashed red lines), showing excellent agreement. The linear theory predictions (dotted grey lines) indicate the amplitude of the non-linear boost $B(k)$ at each redshift.
This agreement can be seen more clearly in Fig.~\ref{fig:aletheia_desi_comparison_ratio}, which presents the ratio 
$\mathcal{E}_P(k)/P(k)$ between the predictions of \texttt{Aletheia} and  the N-body simulation results. The grey shaded regions mark a $\pm 1\%$ agreement band at each redshift. The sub-percent agreement on small scales (where cosmic variance is minimal) shows how the evolution mapping framework provides a powerful method for modelling the non-linear matter power spectrum even in cosmological scenarios far from the fiducial $\Lambda$CDM model.

The evolution mapping approach provides a significant advantage in handling dark energy models, as the framework does not assume a specific parametrisation for the evolution of $w_{\mathrm{DE}}(a)$, such as the standard CPL form \citep{Chevallier2001, Linder2003}. This is possible because the effects of the dark energy only enter the emulator through the background expansion history, which influences $\sigma_{12}$ and the growth history parameter $\tilde{x}$. 
However, this flexibility is limited to models in which dark energy perturbations are negligible and the growth of structure is scale-independent at the linear level. More general models that include dark energy perturbations would require an extension of the current framework.

\subsection{Validation of the resolution correction}
\label{ssec:validation_resolution}

We validated the effectiveness and general applicability of the resolution correction, $\mathcal{C}_{\mathrm{s}}(k, \sigma_{12})$, using an independent high-resolution simulation of the same challenging DESI-motivated cosmology described in Section~\ref{ssec:validation_full_emu}. 
We ran a paired-fixed simulation of this specific cosmology using $2048^3$ particles in a $700\,{\rm Mpc}$ box, and obtained high-resolution power spectrum estimates $P_{\rm HR}(k)$ at six target redshifts ($z=3$ to $z=0$).

\begin{figure}
    \includegraphics[width=0.95\columnwidth]{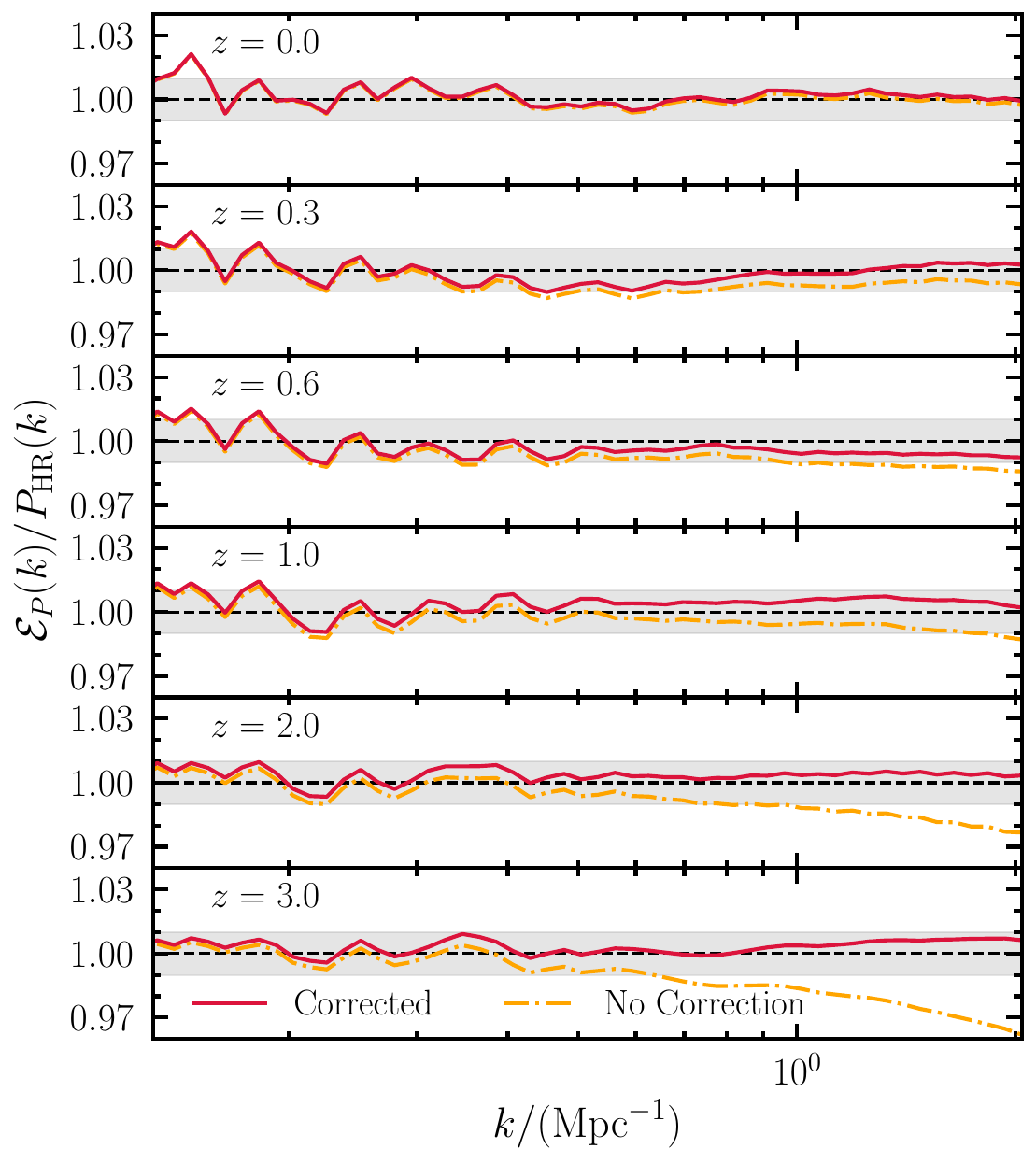}
    \caption{The ratio of the \texttt{Aletheia} predictions to the power spectra inferred from high-resolution simulations at varying  redshifts. The dot-dashed lines show the results of the uncorrected emulator prediction $\mathcal{E}_P(k)$, which underestimates the power at high $k$ for low clustering amplitudes. The solid lines show the results of the final emulator prediction $\mathcal{E}_{\mathrm{final}}(k)$, including the resolution correction described in Secton~\ref{sec:resolution}. The corrected predictions are consistent with $P_{\rm HR}(k)$ across all scales.}
    \label{fig:resolution_validation_plot}
\end{figure}

Figure~\ref{fig:resolution_validation_plot} presents the ratio of the predictions of \texttt{Aletheia} to this high-resolution simulation. Each of the six panels shows the ratio obtained without applying the resolution correction ($\mathcal{E}_P(k)/P_{\mathrm{HR}}(k)$, dot-dashed lines) and the ratio obtained with the correction ($\mathcal{E}_{\mathrm{final}}(k)/P_{\mathrm{HR}}(k)$, solid lines).

As these high-resolution simulations have a smaller volume, the impact of cosmic variance on the measured $P(k)$ is larger, leading to increased scatter and fluctuations in the ratios compared to the tests presented in Section~\ref{ssec:validation_full_emu}. Despite this increased uncertainty, these results demonstrate a clear benefit from the correction. At high redshift, corresponding to low clustering amplitudes, the uncorrected results show a systematic deficit of power at high $k$ with respect to the high-resolution simulation. However, the corrected prediction is consistent with the high-resolution simulation results across the full $k$-range, confirming that $\mathcal{C}_{\mathrm{s}}(k, \sigma_{12})$ successfully accounts for the resolution-induced power suppression, even for 
a cosmology that deviates significantly from the fiducial one used to define the correction.

\section{Conclusions}
\label{sec:conclusions}

In this paper, we have presented {\tt Aletheia}, a new emulator for the non-linear matter power spectrum, $P(k)$ based on the evolution mapping framework proposed by \citet{Sanchez2022}. This approach addresses the limitations of traditional emulation techniques, which struggle with wide parameter space coverage and redshift generalisation, by decoupling cosmological dependencies to offer a highly flexible and accurate analysis tool. The core of this framework relies on focusing on $h$-independent cosmological parameters, which can be classified into those governing the shape of the linear power spectrum ($\bm{\Theta}_{\mathrm{s}}$) and those controlling its subsequent growth ($\bm{\Theta}_{\mathrm{e}}$). The influence of the evolution parameters and redshift can be largely compressed into a single amplitude parameter, $\sigma_{12}$, with small deviations from a universal trend described in terms of the integrated growth history parameter $\tilde{x}$. Building upon this principle, {\tt Aletheia} employs a two-stage emulation strategy: a first emulator reproduces the non-linear boost factor $B(k)$ as a function of $\bm{\Theta}_{\mathrm{s}}$ and $\sigma_{12}$, and a second emulator accounts for deviations from this ideal mapping with a linear correction dependent on $\tilde{x}$. This design allows {\tt Aletheia} to efficiently generate predictions across a broad spectrum of cosmologies and redshifts without the need for dense, explicit sampling of the full parameter space.
Our final emulator predictions include a  correction to account for resolution effects at high wavenumbers, further extending the accuracy of the predictions more deeply into the non-linear regime.

Our validation against N-body simulations demonstrates the high accuracy and robustness of {\tt Aletheia}. The individual emulator components, $\mathcal{E}_B(k)$ and $\mathcal{E}_{\partial R/\partial\tilde{x}}(k)$, were shown to predict their respective quantities with sub-percent accuracy. When combined, the full {\tt Aletheia} emulator provides excellent agreement with N-body simulations across a range of redshifts and for various dynamical dark energy cosmological models. A direct comparison with the state-of-the-art emulator {\tt EuclidEmulator2} showed that, while both give consistent results, {\tt Aletheia} exhibits a significantly lower variance in its predictions (approximately $0.2\%$ compared to approximately $1\%$ for {\tt EuclidEmulator2}). This smaller variance is a direct consequence of the evolution mapping strategy, which drastically reduces the dimensionality of the problem by leveraging the physical degeneracies in structure growth, leading to more stable and precise predictions from a given training set. {\tt Aletheia}'s design is also more general, allowing it to accurately reproduce the power spectrum for the best-fit dynamic dark energy cosmology derived from recent DESI, CMB, and SN data \citep{Adame2025_DESIBAO1}, whose parameters lie well outside the typical training ranges of conventional emulators. Furthermore, our approach can also provide predictions of $P(k)$ for an arbitrary evolution of $w_{\mathrm{DE}}(a)$, as long as these models remain in a regime where dark energy perturbations are negligible and the growth of structure remains scale-independent at the linear level.

The success of {\tt Aletheia} as a proof-of-concept demonstrates the potential of the evolution mapping framework. The emulator is publicly available as a Python package, and a description of its usage is provided in Appendix~\ref{sec:appendix_code}. 
In future work, we plan to extend this setup to emulate other non-linear statistics of the large-scale structure, such as the galaxy-matter cross-power spectrum, the galaxy auto-power spectrum, and higher-order statistics like the bispectrum, further enhancing the capabilities of the {\tt Aletheia} package for cosmological analyses.

Beyond its direct application in building emulators, the evolution mapping framework offers a more physically intuitive understanding of how cosmic structures evolve across different cosmological scenarios. It provides a powerful way to simplify the description of the non-linear evolution of the cosmic density field. The approach fundamentally highlights the importance of expressing theoretical predictions in a manner that is independent of the dimensionless Hubble parameter, $h$, simplifying the apparently complex interplay of cosmological parameters in the formation of the LSS of the Universe.

\section*{Acknowledgements}

We would like to thank Jiamin Hou, Soumadeep Maiti, Andrea Pezzotta, Agne
Semenaite, Mathias Garny, Alex Eggemeier, Mart\'in Crocce, Rom\'an
Scoccimarro, and Lukas Schw\"orer for their help and useful discussions. The AletheiaEmu simulations were carried out and
post-processed on the HPC system Raven of the Max Planck Computing and Data
Facility (MPCDF, \url{https://www.mpcdf.mpg.de}) in Garching, Germany. This research was supported
by the Excellence Cluster ORIGINS, which is funded by the Deutsche
Forschungsgemeinschaft (DFG, German Research Foundation) under Germany's
Excellence Strategy - EXC-2094 - 390783311. JGJ acknowledges funding by ANID (Beca Doctorado Nacional, Folio 21210846). FR
acknowledges support from the ICTP through the Junior Associates Programme
2023-2028. This project has received funding from the European Union’s
HORIZON-MSCA-2021-SE-01 Research and Innovation programme under the Marie
Sklodowska-Curie grant agreement number 101086388 - Project acronym: LACEGAL. 

\section*{Data Availability}

The {\tt Aletheia} emulator described in this work is publicly available and registered on the Python Package Index. The simulation data underlying this article will be shared on reasonable request to the corresponding authors.



\bibliographystyle{mnras}
\bibliography{Aletheia1} 

\begin{thebibliography}{}
\makeatletter
\relax
\def\mn@urlcharsother{\let\do\@makeother \do\$\do\&\do\#\do\^\do\_\do\%\do\~}
\def\mn@doi{\begingroup\mn@urlcharsother \@ifnextchar [ {\mn@doi@} {\mn@doi@[]}}
\def\mn@doi@[#1]#2{\def\@tempa{#1}\ifx\@tempa\@empty \href {http://dx.doi.org/#2} {doi:#2}\else \href {http://dx.doi.org/#2} {#1}\fi \endgroup}
\def\mn@eprint#1#2{\mn@eprint@#1:#2::\@nil}
\def\mn@eprint@arXiv#1{\href {http://arxiv.org/abs/#1} {{\tt arXiv:#1}}}
\def\mn@eprint@dblp#1{\href {http://dblp.uni-trier.de/rec/bibtex/#1.xml} {dblp:#1}}
\def\mn@eprint@#1:#2:#3:#4\@nil{\def\@tempa {#1}\def\@tempb {#2}\def\@tempc {#3}\ifx \@tempc \@empty \let \@tempc \@tempb \let \@tempb \@tempa \fi \ifx \@tempb \@empty \def\@tempb {arXiv}\fi \@ifundefined {mn@eprint@\@tempb}{\@tempb:\@tempc}{\expandafter \expandafter \csname mn@eprint@\@tempb\endcsname \expandafter{\@tempc}}}

\bibitem[\protect\citeauthoryear{{Abbott} et~al.,}{{Abbott} et~al.}{2022}]{Abbott2022}
{Abbott} T.~M.~C.,  et~al., 2022, \mn@doi [\prd] {10.1103/PhysRevD.105.023520}, \href {https://ui.adsabs.harvard.edu/abs/2022PhRvD.105b3520A} {105, 023520}

\bibitem[\protect\citeauthoryear{Alam et~al.,}{Alam et~al.}{2017}]{Alam2017}
Alam S.,  et~al., 2017, \mn@doi [\mnras] {10.1093/mnras/stx721}, 470, 2617–2652

\bibitem[\protect\citeauthoryear{{Alam} et~al.,}{{Alam} et~al.}{2021}]{Alam2021}
{Alam} S.,  et~al., 2021, \mn@doi [\prd] {10.1103/PhysRevD.103.083533}, \href {https://ui.adsabs.harvard.edu/abs/2021PhRvD.103h3533A} {103, 083533}

\bibitem[\protect\citeauthoryear{{Anderson} et~al.,}{{Anderson} et~al.}{2014}]{Anderson2014}
{Anderson} L.,  et~al., 2014, \mn@doi [\mnras] {10.1093/mnras/stu523}, \href {http://adsabs.harvard.edu/abs/2014MNRAS.441...24A} {441, 24}

\bibitem[\protect\citeauthoryear{{Angulo} \& {Pontzen}}{{Angulo} \& {Pontzen}}{2016}]{Angulo2016}
{Angulo} R.~E.,  {Pontzen} A.,  2016, \mn@doi [\mnras] {10.1093/mnrasl/slw098}, \href {https://ui.adsabs.harvard.edu/abs/2016MNRAS.462L...1A} {462, L1}

\bibitem[\protect\citeauthoryear{{Angulo}, {Zennaro}, {Contreras}, {Aric{\'o}}, {Pellejero-Iba{\~n}ez}  \& {St{\"u}cker}}{{Angulo} et~al.}{2021}]{Angulo2021_BACCO}
{Angulo} R.~E.,  {Zennaro} M.,  {Contreras} S.,  {Aric{\'o}} G.,  {Pellejero-Iba{\~n}ez} M.,   {St{\"u}cker} J.,  2021, \mn@doi [\mnras] {10.1093/mnras/stab2018}, \href {https://ui.adsabs.harvard.edu/abs/2021MNRAS.tmp.1931A} {}

\bibitem[\protect\citeauthoryear{{Bocquet}, {Heitmann}, {Habib}, {Lawrence}, {Uram}, {Frontiere}, {Pope}  \& {Finkel}}{{Bocquet} et~al.}{2020}]{Bocquet2020}
{Bocquet} S.,  {Heitmann} K.,  {Habib} S.,  {Lawrence} E.,  {Uram} T.,  {Frontiere} N.,  {Pope} A.,   {Finkel} H.,  2020, \mn@doi [\apj] {10.3847/1538-4357/abac5c}, \href {https://ui.adsabs.harvard.edu/abs/2020ApJ...901....5B} {901, 5}

\bibitem[\protect\citeauthoryear{{Brout} et~al.,}{{Brout} et~al.}{2022}]{Brout2022_PantheonPlus}
{Brout} D.,  et~al., 2022, \mn@doi [\apj] {10.3847/1538-4357/ac8e04}, \href {https://ui.adsabs.harvard.edu/abs/2022ApJ...938..110B} {938, 110}

\bibitem[\protect\citeauthoryear{{Chevallier} \& {Polarski}}{{Chevallier} \& {Polarski}}{2001}]{Chevallier2001}
{Chevallier} M.,  {Polarski} D.,  2001, \mn@doi [International Journal of Modern Physics D] {10.1142/S0218271801000822}, \href {http://adsabs.harvard.edu/abs/2001IJMPD..10..213C} {10, 213}

\bibitem[\protect\citeauthoryear{Cole et~al.,}{Cole et~al.}{2005}]{Cole2005}
Cole S.,  et~al., 2005, \mn@doi [\mnras] {10.1111/j.1365-2966.2005.09318.x}, 362, 505–534

\bibitem[\protect\citeauthoryear{{Crocce}, {Pueblas}  \& {Scoccimarro}}{{Crocce} et~al.}{2006}]{crocce_2lptic}
{Crocce} M.,  {Pueblas} S.,   {Scoccimarro} R.,  2006, \mn@doi [\mnras] {10.1111/j.1365-2966.2006.11040.x}, \href {https://ui.adsabs.harvard.edu/abs/2006MNRAS.373..369C} {373, 369}

\bibitem[\protect\citeauthoryear{{Crocce}, {Pueblas}  \& {Scoccimarro}}{{Crocce} et~al.}{2012}]{Crocce2012_code2lptic}
{Crocce} M.,  {Pueblas} S.,   {Scoccimarro} R.,  2012, {2LPTIC: 2nd-order Lagrangian Perturbation Theory Initial Conditions}, Astrophysics Source Code Library, record ascl:1201.005

\bibitem[\protect\citeauthoryear{{DESI Collaboration} et~al.,}{{DESI Collaboration} et~al.}{2025a}]{DESI2025_BAO2}
{DESI Collaboration} et~al., 2025a, \mn@doi [arXiv e-prints] {10.48550/arXiv.2503.14738}, \href {https://ui.adsabs.harvard.edu/abs/2025arXiv250314738D} {p. arXiv:2503.14738}

\bibitem[\protect\citeauthoryear{{DESI Collaboration} et~al.,}{{DESI Collaboration} et~al.}{2025b}]{Adame2025_DESIBAO1}
{DESI Collaboration} et~al., 2025b, \mn@doi [\jcap] {10.1088/1475-7516/2025/02/021}, \href {https://ui.adsabs.harvard.edu/abs/2025JCAP...02..021A} {2025, 021}

\bibitem[\protect\citeauthoryear{{DESI Collaboration} et~al.,}{{DESI Collaboration} et~al.}{2025c}]{DESI2024_FS1}
{DESI Collaboration} et~al., 2025c, \mn@doi [\jcap] {10.1088/1475-7516/2025/07/028}, \href {https://ui.adsabs.harvard.edu/abs/2025JCAP...07..028A} {2025, 028}

\bibitem[\protect\citeauthoryear{{Dalal} et~al.,}{{Dalal} et~al.}{2023}]{Dalal2023}
{Dalal} R.,  et~al., 2023, \mn@doi [\prd] {10.1103/PhysRevD.108.123519}, \href {https://ui.adsabs.harvard.edu/abs/2023PhRvD.108l3519D} {108, 123519}

\bibitem[\protect\citeauthoryear{{DeRose} et~al.,}{{DeRose} et~al.}{2019}]{AemulusI}
{DeRose} J.,  et~al., 2019, \mn@doi [\apj] {10.3847/1538-4357/ab1085}, \href {https://ui.adsabs.harvard.edu/abs/2019ApJ...875...69D} {875, 69}

\bibitem[\protect\citeauthoryear{{Eggemeier}, {Camacho-Quevedo}, {Pezzotta}, {Crocce}, {Scoccimarro}  \& {S{\'a}nchez}}{{Eggemeier} et~al.}{2023}]{Eggemeier2023}
{Eggemeier} A.,  {Camacho-Quevedo} B.,  {Pezzotta} A.,  {Crocce} M.,  {Scoccimarro} R.,   {S{\'a}nchez} A.~G.,  2023, \mn@doi [\mnras] {10.1093/mnras/stac3667}, \href {https://ui.adsabs.harvard.edu/abs/2023MNRAS.519.2962E} {519, 2962}

\bibitem[\protect\citeauthoryear{{Eisenstein} et~al.,}{{Eisenstein} et~al.}{2005}]{Eisenstein2005}
{Eisenstein} D.~J.,  et~al., 2005, \mn@doi [\apj] {10.1086/466512}, \href {https://ui.adsabs.harvard.edu/abs/2005ApJ...633..560E} {633, 560}

\bibitem[\protect\citeauthoryear{{Esposito}, {S{\'a}nchez}, {Bel}  \& {Ruiz}}{{Esposito} et~al.}{2024}]{Esposito2024_VelEvoMap}
{Esposito} M.,  {S{\'a}nchez} A.~G.,  {Bel} J.,   {Ruiz} A.~N.,  2024, \mn@doi [\mnras] {10.1093/mnras/stae2351}, \href {https://ui.adsabs.harvard.edu/abs/2024MNRAS.534.3906E} {534, 3906}

\bibitem[\protect\citeauthoryear{{Euclid Collaboration} et~al.,}{{Euclid Collaboration} et~al.}{2019}]{Euclidemulator}
{Euclid Collaboration} et~al., 2019, \mn@doi [\mnras] {10.1093/mnras/stz197}, \href {https://ui.adsabs.harvard.edu/abs/2019MNRAS.484.5509E} {484, 5509}

\bibitem[\protect\citeauthoryear{{Euclid Collaboration} et~al.,}{{Euclid Collaboration} et~al.}{2021}]{EuclidEmulator2}
{Euclid Collaboration} et~al., 2021, \mn@doi [\mnras] {10.1093/mnras/stab1366}, \href {https://ui.adsabs.harvard.edu/abs/2021MNRAS.505.2840E} {505, 2840}

\bibitem[\protect\citeauthoryear{{Euclid Collaboration} et~al.,}{{Euclid Collaboration} et~al.}{2025}]{Euclid_overview}
{Euclid Collaboration} et~al., 2025, \mn@doi [\aap] {10.1051/0004-6361/202450810}, \href {https://ui.adsabs.harvard.edu/abs/2025A&A...697A...1E} {697, A1}

\bibitem[\protect\citeauthoryear{{Garny} \& {Taule}}{{Garny} \& {Taule}}{2021}]{Garny2021}
{Garny} M.,  {Taule} P.,  2021, \mn@doi [\jcap] {10.1088/1475-7516/2021/01/020}, \href {https://ui.adsabs.harvard.edu/abs/2021JCAP...01..020G} {2021, 020}

\bibitem[\protect\citeauthoryear{{Garrison}, {Eisenstein}, {Ferrer}, {Tinker}, {Pinto}  \& {Weinberg}}{{Garrison} et~al.}{2018}]{Garrison2018}
{Garrison} L.~H.,  {Eisenstein} D.~J.,  {Ferrer} D.,  {Tinker} J.~L.,  {Pinto} P.~A.,   {Weinberg} D.~H.,  2018, \mn@doi [The Astrophysical Journal Supplement Series] {10.3847/1538-4365/aabfd3}, \href {https://ui.adsabs.harvard.edu/abs/2018ApJS..236...43G} {236, 43}

\bibitem[\protect\citeauthoryear{{Grieb} et~al.,}{{Grieb} et~al.}{2017}]{Grieb2017}
{Grieb} J.~N.,  et~al., 2017, \mn@doi [\mnras] {10.1093/mnras/stw3384}, \href {http://adsabs.harvard.edu/abs/2017MNRAS.467.2085G} {467, 2085}

\bibitem[\protect\citeauthoryear{{Hamann}, {Hannestad}, {Lesgourgues}, {Rampf}  \& {Wong}}{{Hamann} et~al.}{2010}]{Hamann2010}
{Hamann} J.,  {Hannestad} S.,  {Lesgourgues} J.,  {Rampf} C.,   {Wong} Y. Y.~Y.,  2010, \mn@doi [\jcap] {10.1088/1475-7516/2010/07/022}, \href {https://ui.adsabs.harvard.edu/abs/2010JCAP...07..022H} {2010, 022}

\bibitem[\protect\citeauthoryear{{Hamilton}, {Kumar}, {Lu}  \& {Matthews}}{{Hamilton} et~al.}{1991}]{Hamilton1991}
{Hamilton} A.~J.~S.,  {Kumar} P.,  {Lu} E.,   {Matthews} A.,  1991, \mn@doi [\apjl] {10.1086/186057}, \href {https://ui.adsabs.harvard.edu/abs/1991ApJ...374L...1H} {374, L1}

\bibitem[\protect\citeauthoryear{{Heitmann}, {White}, {Wagner}, {Habib}  \& {Higdon}}{{Heitmann} et~al.}{2010}]{coyote2010}
{Heitmann} K.,  {White} M.,  {Wagner} C.,  {Habib} S.,   {Higdon} D.,  2010, \mn@doi [\apj] {10.1088/0004-637X/715/1/104}, \href {https://ui.adsabs.harvard.edu/abs/2010ApJ...715..104H} {715, 104}

\bibitem[\protect\citeauthoryear{{Heitmann} et~al.,}{{Heitmann} et~al.}{2016}]{Heitman2016}
{Heitmann} K.,  et~al., 2016, \mn@doi [\apj] {10.3847/0004-637X/820/2/108}, \href {https://ui.adsabs.harvard.edu/abs/2016ApJ...820..108H} {820, 108}

\bibitem[\protect\citeauthoryear{Hockney \& Eastwood}{Hockney \& Eastwood}{1988}]{Hockney1988}
Hockney R.~W.,  Eastwood J.~W.,  1988, Computer Simulation Using Particles, 1 edn.
CRC Press, Boca Raton, FL, \mn@doi{10.1201/9780367806934}, \url {https://doi.org/10.1201/9780367806934}

\bibitem[\protect\citeauthoryear{{Ivanov}, {Simonovi{\'c}}  \& {Zaldarriaga}}{{Ivanov} et~al.}{2020}]{Ivanov2020}
{Ivanov} M.~M.,  {Simonovi{\'c}} M.,   {Zaldarriaga} M.,  2020, \mn@doi [\jcap] {10.1088/1475-7516/2020/05/042}, \href {https://ui.adsabs.harvard.edu/abs/2020JCAP...05..042I} {2020, 042}

\bibitem[\protect\citeauthoryear{{Li} et~al.,}{{Li} et~al.}{2023}]{Li2023}
{Li} X.,  et~al., 2023, \mn@doi [\prd] {10.1103/PhysRevD.108.123518}, \href {https://ui.adsabs.harvard.edu/abs/2023PhRvD.108l3518L} {108, 123518}

\bibitem[\protect\citeauthoryear{{Linder}}{{Linder}}{2003}]{Linder2003}
{Linder} E.~V.,  2003, \mn@doi [Physical Review Letters] {10.1103/PhysRevLett.90.091301}, \href {http://adsabs.harvard.edu/abs/2003PhRvL..90i1301L} {90, 091301}

\bibitem[\protect\citeauthoryear{{Louis} et~al.,}{{Louis} et~al.}{2025}]{Thibaut2025_ACT}
{Louis} T.,  et~al., 2025, \mn@doi [arXiv e-prints] {10.48550/arXiv.2503.14452}, \href {https://ui.adsabs.harvard.edu/abs/2025arXiv250314452L} {p. arXiv:2503.14452}

\bibitem[\protect\citeauthoryear{{McKay}, {Beckman}  \& {Conover}}{{McKay} et~al.}{1979}]{mckay1979_LH}
{McKay} M.~D.,  {Beckman} R.~J.,   {Conover} W.~J.,  1979, Technometrics, 21, 239

\bibitem[\protect\citeauthoryear{{Peacock} \& {Dodds}}{{Peacock} \& {Dodds}}{1996}]{Peacock1996}
{Peacock} J.~A.,  {Dodds} S.~J.,  1996, \mn@doi [\mnras] {10.1093/mnras/280.3.L19}, \href {https://ui.adsabs.harvard.edu/abs/1996MNRAS.280L..19P} {280, L19}

\bibitem[\protect\citeauthoryear{{Pedregosa} et~al.,}{{Pedregosa} et~al.}{2011}]{Pedregosa2011_scikitlearn}
{Pedregosa} F.,  et~al., 2011, \mn@doi [Journal of Machine Learning Research] {10.48550/arXiv.1201.0490}, \href {https://ui.adsabs.harvard.edu/abs/2011JMLR...12.2825P} {12, 2825}

\bibitem[\protect\citeauthoryear{{Pezzotta} et~al.,}{{Pezzotta} et~al.}{2025}]{Pezzotta2025_NuEvoMap}
{Pezzotta} A.,  et~al., 2025, \mn@doi [\prd] {10.1103/vy3h-p92n}, \href {https://ui.adsabs.harvard.edu/abs/2025PhRvD.112b3520P} {112, 023520}

\bibitem[\protect\citeauthoryear{{Planck Collaboration} et~al.,}{{Planck Collaboration} et~al.}{2020}]{Planck2018}
{Planck Collaboration} et~al., 2020, \mn@doi [\aap] {10.1051/0004-6361/201833910}, \href {https://ui.adsabs.harvard.edu/abs/2020A&A...641A...6P} {641, A6}

\bibitem[\protect\citeauthoryear{{Rasmussen} \& {Williams}}{{Rasmussen} \& {Williams}}{2006}]{Rasmussen2006_GPML}
{Rasmussen} C.~E.,  {Williams} C. K.~I.,  2006, {Gaussian Processes for Machine Learning}.
{The MIT Press}

\bibitem[\protect\citeauthoryear{{S{\'a}nchez}}{{S{\'a}nchez}}{2020}]{Sanchez2020}
{S{\'a}nchez} A.~G.,  2020, \mn@doi [\prd] {10.1103/PhysRevD.102.123511}, \href {https://ui.adsabs.harvard.edu/abs/2020PhRvD.102l3511S} {102, 123511}

\bibitem[\protect\citeauthoryear{{S{\'a}nchez}, {Baugh}, {Percival}, {Peacock}, {Padilla}, {Cole}, {Frenk}  \& {Norberg}}{{S{\'a}nchez} et~al.}{2006}]{Sanchez2006}
{S{\'a}nchez} A.~G.,  {Baugh} C.~M.,  {Percival} W.~J.,  {Peacock} J.~A.,  {Padilla} N.~D.,  {Cole} S.,  {Frenk} C.~S.,   {Norberg} P.,  2006, \mn@doi [\mnras] {10.1111/j.1365-2966.2005.09833.x}, \href {https://ui.adsabs.harvard.edu/abs/2006MNRAS.366..189S} {366, 189}

\bibitem[\protect\citeauthoryear{{S{\'a}nchez} et~al.,}{{S{\'a}nchez} et~al.}{2017}]{Sanchez2017b}
{S{\'a}nchez} A.~G.,  et~al., 2017, \mn@doi [\mnras] {10.1093/mnras/stw2495}, \href {http://adsabs.harvard.edu/abs/2017MNRAS.464.1493S} {464, 1493}

\bibitem[\protect\citeauthoryear{{S{\'a}nchez}, {Ruiz}, {Jara}  \& {Padilla}}{{S{\'a}nchez} et~al.}{2022}]{Sanchez2022}
{S{\'a}nchez} A.~G.,  {Ruiz} A.~N.,  {Jara} J.~G.,   {Padilla} N.~D.,  2022, \mn@doi [\mnras] {10.1093/mnras/stac1656}, \href {https://ui.adsabs.harvard.edu/abs/2022MNRAS.514.5673S} {514, 5673}

\bibitem[\protect\citeauthoryear{{Scoccimarro}, {Colombi}, {Fry}, {Frieman}, {Hivon}  \& {Melott}}{{Scoccimarro} et~al.}{1998}]{Scoccimarro1998}
{Scoccimarro} R.,  {Colombi} S.,  {Fry} J.~N.,  {Frieman} J.~A.,  {Hivon} E.,   {Melott} A.,  1998, \mn@doi [\apj] {10.1086/305399}, \href {https://ui.adsabs.harvard.edu/abs/1998ApJ...496..586S} {496, 586}

\bibitem[\protect\citeauthoryear{{Scolnic} et~al.,}{{Scolnic} et~al.}{2018}]{Scolnic2018_Pantheon}
{Scolnic} D.~M.,  et~al., 2018, \mn@doi [\apj] {10.3847/1538-4357/aab9bb}, \href {https://ui.adsabs.harvard.edu/abs/2018ApJ...859..101S} {859, 101}

\bibitem[\protect\citeauthoryear{{Scolnic} et~al.,}{{Scolnic} et~al.}{2022}]{Scolnic2022_PantheonPlus}
{Scolnic} D.,  et~al., 2022, \mn@doi [\apj] {10.3847/1538-4357/ac8b7a}, \href {https://ui.adsabs.harvard.edu/abs/2022ApJ...938..113S} {938, 113}

\bibitem[\protect\citeauthoryear{{Sefusatti}, {Crocce}, {Scoccimarro}  \& {Couchman}}{{Sefusatti} et~al.}{2016}]{Sefusatti2016}
{Sefusatti} E.,  {Crocce} M.,  {Scoccimarro} R.,   {Couchman} H.~M.~P.,  2016, \mn@doi [\mnras] {10.1093/mnras/stw1229}, \href {https://ui.adsabs.harvard.edu/abs/2016MNRAS.460.3624S} {460, 3624}

\bibitem[\protect\citeauthoryear{{Smith} et~al.,}{{Smith} et~al.}{2003}]{Smith2003}
{Smith} R.~E.,  et~al., 2003, \mn@doi [\mnras] {10.1046/j.1365-8711.2003.06503.x}, \href {https://ui.adsabs.harvard.edu/abs/2003MNRAS.341.1311S} {341, 1311}

\bibitem[\protect\citeauthoryear{{Springel}, {Pakmor}, {Zier}  \& {Reinecke}}{{Springel} et~al.}{2021}]{Springel2021_Gadget4}
{Springel} V.,  {Pakmor} R.,  {Zier} O.,   {Reinecke} M.,  2021, \mn@doi [\mnras] {10.1093/mnras/stab1855}, \href {https://ui.adsabs.harvard.edu/abs/2021MNRAS.506.2871S} {506, 2871}

\bibitem[\protect\citeauthoryear{{Stein}}{{Stein}}{1987}]{stein1987_MLH}
{Stein} M.,  1987, Technometrics, 29, 143151

\bibitem[\protect\citeauthoryear{{Takahashi}}{{Takahashi}}{2008}]{Takahashi2008}
{Takahashi} R.,  2008, \mn@doi [Progress of Theoretical Physics] {10.1143/PTP.120.549}, \href {https://ui.adsabs.harvard.edu/abs/2008PThPh.120..549T} {120, 549}

\bibitem[\protect\citeauthoryear{{Taruya}}{{Taruya}}{2016}]{Taruya2016}
{Taruya} A.,  2016, \mn@doi [\prd] {10.1103/PhysRevD.94.023504}, \href {https://ui.adsabs.harvard.edu/abs/2016PhRvD..94b3504T} {94, 023504}

\bibitem[\protect\citeauthoryear{{Tristram} et~al.,}{{Tristram} et~al.}{2024}]{Planck2024}
{Tristram} M.,  et~al., 2024, \mn@doi [\aap] {10.1051/0004-6361/202348015}, \href {https://ui.adsabs.harvard.edu/abs/2024A&A...682A..37T} {682, A37}

\bibitem[\protect\citeauthoryear{{Tr{\"o}ster} et~al.,}{{Tr{\"o}ster} et~al.}{2020}]{Troster2020}
{Tr{\"o}ster} T.,  et~al., 2020, \mn@doi [\aap] {10.1051/0004-6361/201936772}, \href {https://ui.adsabs.harvard.edu/abs/2020A&A...633L..10T} {633, L10}

\bibitem[\protect\citeauthoryear{{Wright} et~al.,}{{Wright} et~al.}{2025}]{Wright2025}
{Wright} A.~H.,  et~al., 2025, \mn@doi [arXiv e-prints] {10.48550/arXiv.2503.19441}, \href {https://ui.adsabs.harvard.edu/abs/2025arXiv250319441W} {p. arXiv:2503.19441}

\bibitem[\protect\citeauthoryear{{Yang}, {Bird}  \& {Ho}}{{Yang} et~al.}{2025}]{Yang2025}
{Yang} Y.,  {Bird} S.,   {Ho} M.-F.,  2025, \mn@doi [\prd] {10.1103/PhysRevD.111.083529}, \href {https://ui.adsabs.harvard.edu/abs/2025PhRvD.111h3529Y} {111, 083529}

\bibitem[\protect\citeauthoryear{{d'Amico}, {Gleyzes}, {Kokron}, {Markovic}, {Senatore}, {Zhang}, {Beutler}  \& {Gil-Mar{\'\i}n}}{{d'Amico} et~al.}{2020}]{Damico2020}
{d'Amico} G.,  {Gleyzes} J.,  {Kokron} N.,  {Markovic} K.,  {Senatore} L.,  {Zhang} P.,  {Beutler} F.,   {Gil-Mar{\'\i}n} H.,  2020, \mn@doi [\jcap] {10.1088/1475-7516/2020/05/005}, \href {https://ui.adsabs.harvard.edu/abs/2020JCAP...05..005D} {2020, 005}

\makeatother
\end{thebibliography}



\appendix

\section{Functionality of the {\tt Aletheia} package}
\label{sec:appendix_code}

The non-linear power spectrum emulator described in this work is publicly available as a Python package named \texttt{Aletheia}. This appendix provides a brief guide to its installation and basic usage. For more detailed information, tutorials, and the source code, we refer the reader to the public repository\footnote{\url{https://gitlab.mpcdf.mpg.de/arielsan/aletheia}} and the full online documentation\footnote{\url{https://aletheia-46606f.pages.mpcdf.de/index.html}}.

The package is registered on the Python Package Index (PyPI) and can be installed in a Python environment using \texttt{pip}, as
\begin{verbatim}
pip install AletheiaCosmo
\end{verbatim}
This command will automatically handle the installation of the required dependencies, which include \texttt{numpy}, \texttt{scipy}, and \texttt{camb}.

The core of the package is the \texttt{AletheiaEmu} class. The typical workflow involves creating a cosmology dictionary, initialising the emulator, and then calling its main prediction method. A minimal working example is shown below.

\begin{verbatim}
import numpy as np
from aletheiacosmo import AletheiaEmu

# 1. Define the cosmology using the helper function.
# This function handles the conversion of parameters
# and ensures the dictionary is correctly formatted
# for the emulator.
cosmo_params = AletheiaEmu.create_cosmo_dict(
    h=0.67,
    omega_b=0.0224,
    omega_c=0.120,
    n_s=0.96,
    A_s=2.1e-9,
    model='LCDM'  # Automatically sets w0=-1, wa=0
)

# 2. Initialise the emulator instance.
# This loads the internal GP models 
# and correction functions.
emu = AletheiaEmu()

# 3. Define the wavenumbers and redshift for the 
# prediction.
k_values = np.logspace(-2, 0.3, 200)  # k in 1/Mpc
redshift = 1.0

# 4. Get the non-linear power spectrum.
# The get_pnl method returns a NumPy array 
# of P_NL(k).
P_NL = emu.get_pnl(k_values, cosmo_params, redshift)

# The result can now be used for analysis.
\end{verbatim}

The primary method of the class is \texttt{get\_pnl(k, cospar, z)}, which takes an array of wavenumbers in units of ${\rm Mpc}^{-1}$, a cosmology dictionary, and a single redshift, and returns the non-linear power spectrum in units of ${\rm Mpc}^3$. The package also includes internal validation to ensure that the input cosmology lies within the parameter space where the emulator was trained, raising an error if the boundaries are exceeded.


\bsp	
\label{lastpage}
\end{document}